\documentclass[dvips,12pt,a4paper]{article}
\newcommand{\Section}[1]{\section{#1}\setcounter{equation}{0}}

\usepackage{fleqn,amsfonts,amssymb,amsmath,graphicx,version,theorem,epsfig}

\newcommand{\Z}{\mathbb Z}
\newcommand{\N}{\mathbb N}
\newcommand{\C}{\mathbb C}

\newcommand{\R}{\mathbb R}
\newcommand{\Q}{\mathbb Q}

\newcommand{\B}{\cal B}

\newtheorem{theorem}{Theorem}[section]
\newtheorem{assumptio!n}{Assumption}
\newtheorem {lemma}[theorem]{Lemma}
\newtheorem {proposition}[theorem]{Proposition}
\newtheorem {corollary}[theorem]{Corollary}
\theorembodyfont{\normalfont}
\newtheorem {definition}[theorem]{Definition}
\newtheorem {defi}[theorem]{Definition}

\newtheorem {example}[theorem]{Example}

\newtheorem {remark}[theorem]{Remark}

\def\be{\begin{equation}}
\def\ee{\end{equation}}

\newcommand{\beq}{\begin{equation}}
\newcommand{\eeq}{\end{equation}}
\newcommand{\Leq}[1]{\label{#1}\end{equation}}
\newcommand{\beqn}{\begin{eqnarray}}
\newcommand{\eeqn}{\end{eqnarray}}
\newcommand{\beqno}{\begin{eqnarray*}}
\newcommand{\eeqno}{\end{eqnarray*}}

\renewcommand {\l}{\left}
\newcommand{\ri}{\right}
\newcommand {\vep}{\varepsilon}

\newcommand {\vv}{\varphi}

\newcommand {\LA}{\left\langle}
\newcommand {\RA}{\right\rangle}

\newcommand {\eh}{{\textstyle \frac{1}{2}}}

\newcommand {\ar}{\rightarrow}
\newcommand {\sign}{{\rm sign}}

\renewcommand  {\Re}{{\rm Re}}

\newcommand {\tr}{{\rm tr}}

\newcommand {\bC}{{\mathbb C}}

\newcommand {\bN}{{\mathbb N}}
\newcommand {\bR}{{\mathbb R}}
\newcommand {\bZ}{{\mathbb Z}}

\newcommand{\idty}{{\rm 1\mskip-4mu l}} 
\newcommand{\cP}{{\cal P}}

\newcommand{\ov}{\overline}
\newcommand{\bem}{\l(\! \begin{array}}
\newcommand{\eem}{\end{array}\!\ri)}
\newcommand{\bsm}{\left(\begin{smallmatrix}} 
\newcommand{\esm}{\end{smallmatrix}\right)}  
\newcommand{\NN}{\nonumber}

\newcommand{\hp}{{\hat{p}}}

\newcommand{\hq}{{\hat{q}}}

\newcommand{\qmbox}[1]{\quad\mbox{#1}\quad}

\renewcommand {\max}{{{\rm max}}}

\newcommand{\supp}{{\rm supp}}

\newcommand{\northeast}
{\raisebox{1.5mm}{$\lrcorner$}\raisebox{-2mm}{$\ulcorner$}}

\newcommand{\tree}{{\cal T}}
\newcommand{\Pqr}{{\cal P}_{s,r}}

\newcommand {\Gb}{{\bf G}}

\newcommand {\pc}{{p^c\!\!}}
\newcommand {\si}{\sigma}
\newcommand {\qc}{{q^c\!\!}}
\newcommand {\G}{{\bf G}}
\newcommand {\Gk}{{\G}_k}

\newcommand{\Zc}{Z_n^{C}}
\newcommand{\Zg}{Z_n^{G}}
\newcommand{\Psr}{{\cal P}_{s,r}}

\begin{document}
\title{Generalized Farey Trees, Transfer Operators\\ and Phase
Transitions}

\author{Mirko Degli Esposti
\thanks{
   Dipartimento di Matematica, Universit\`a di Bologna,
   Piazza di Porta S. Donato, 5, I-40127 Bologna, Italy,
   e-mail: desposti@dm.unibo.it}
   \and
   Stefano Isola
   \thanks{Dipartimento di Matematica e Informatica, Universit\`a
   di Camerino, via Madonna delle Carceri, 62032 Camerino,
   Italy. e-mail: stefano.isola@unicam.
it.}
\and
   Andreas Knauf
   \thanks{Mathematisches Institut der
   Universit\"at Erlangen-N\"urnberg. Bismarckstr. 1 1/2,
   D-91054 Erlangen, Germany.
   e-mail: knauf@mi.uni-erlangen.de }}
\date{June 1, 2006}
\maketitle

\abstract

\noindent
We consider a family of Markov maps on the unit interval,
interpolating between the tent map and the Farey map. The latter map
is not uniformly expanding. Each map being composed of two
fractional linear transformations, the family generalizes many
particular properties which for the case of the Farey map have been
successfully exploited in number theory. We analyze the dynamics
through the spectral analysis of generalized transfer operators.
Application of the thermodynamic formalism to the family reveals
first and second order phase transitions and unusual properties like
positivity of the interaction function.

\vskip 0.2cm \noindent
{\bf Key words:} Transfer operators, $\zeta$
and partition functions, trees, phase transitions
 \vskip 0.5cm
 \tableofcontents
 \Section{Introduction}
%
The piecewise real-analytic map
\[F_1:[0,1]\to[0,1]\qmbox{,} x\mapsto\l\{\begin{array}{ccc}
\frac{x}{1-x}&,&0\le x\le1/2\\ \frac{1-x}{x}&,&1/2<x\le1\end{array}\ri.\]
is known as the {\em Farey map}.

\noindent  From the ergodic point of view it is of interest, since it is
expanding everywhere but at the fixed point $x=0$ where it has
slope one. This makes this map a simple model of the physical
phenomenon of {\sl intermittency} \cite{PM}. 

\noindent  From the point of
view of number theory, $F_1$ encodes the continued fraction
algorithm as well as the Riemann zeta function. In particular it
has an induced version given by the celebrated Gauss map
\cite{Ma}. 

\noindent In addition, several models of statistical mechanics
have been considered in recent years in connection to Farey
fractions and continued fractions 
\cite{Kn2,Kn3,KO,FKO,LR}. 

\noindent Altogether this
motivates a precise analysis of the dynamics induced by $F_1$. An
effective tool in this analysis is provided by the {\em transfer
operator} associated to the map (see \cite{Ba} for an overview).
For the map $F_1$ the spectrum of the transfer operator when
acting on a suitable space of analytic functions has been studied
in \cite{Is} and \cite{Pr} and turns out to have a continuous
component, in particular no spectral gap. As a consequence, the
Farey map is ergodic w.r.t. the a.c. infinite measure
$\frac{dx}{x}$. Another interesting ergodic invariant measure for
$F_1$ is the Minkowski probability measure $d?$ (for the question
mark function) which is singular w.r.t. Lebesgue measure and turns
out to be the measure of maximal entropy for $F_1$.

\noindent On the other hand, the Minkowski question mark function
conjugates $F_1$ with the much simpler {\em tent map}
\[F_0:[0,1]\to[0,1] \qmbox{,} x\mapsto\l\{\begin{array}{ccc}2x&,&0\le x\le1/2\\ 2(1-x)&,&1/2<x\le1 \end{array}\ri.\]
which is ergodic w.r.t.\ Lebesgue measure and is -- from the point of
view of number theory -- connected with the base $2$ expansion.

\noindent In \cite{GI} it was first noticed that $F_0$ and $F_1$ can
be viewed as instances of a one-parameter family of interval
expanding maps $F_r$ which are continuous, real-analytic and {\em
accessible}, being composed of two M\"obius transformations.

\noindent The present article now follows this line of research and
connects it with aspects coming from thermodynamic formalism.

\noindent  In particular, one aim is to extend the theory developed
in \cite{Kn2} and \cite{Kn3} to a more general class of trees
encoding the dynamics of the maps $F_r$. It turns out that all these
models share several interesting algebraic relations (see below).

\noindent  Besides that, the value of this approach consists in the
fact that some delicate properties of the arithmetic case $r=1$ can
be approached by first studying the corresponding properties
 for $r<1$ by means of spectral techniques, taking advantage of the
 existence of a spectral gap for the transfer operator, and then
 taking the limit $r\to 1$.

\noindent The paper is organized as follows. In {\em Section}
\ref{family}  we introduce the one-parameter family of interval maps
$F_r$ and recall some of its basic properties obtained in \cite{GI}.
We then describe some relevant features of the Ruelle transfer
operators associated to this family including its action on a
suitable invariant Hilbert space of analytic functions.

\noindent In {\em Section} \ref{alberi} we describe how using the
maps $F_r$ we can construct a one-parameter family of binary trees
$\tree (r)$ interpolating between the dyadic tree and the Farey
tree. As the parameter $r$ ranges in the unit interval, the natural
coding associated to real numbers in $[0,1]$ by each tree induces a
H\"older continuous conjugation between the maps $F_r$ and $F_0$,
thus generalizing the Minkowski question mark function (Lemma
\ref{code}). Moreover we show (Proposition \ref{fareysum}) that
$\tree (r)$ can be also constructed by a local generating rule which
generalizes the mediant operation used to generate Farey fractions.

\noindent
A closed expressions for the
trace of the iterates of the transfer operator  in terms of weighted sums over the leaves of the tree is obtained in
Theorem \ref{trace}, along with some consequences both on dynamical zeta functions and Fredholm determinants for the
family $F_r$.

\noindent The trees $\tree (r)$ turn out to be fundamental objects
in establishing a direct connection between the transfer operators
mentioned above and the partition functions of class of spin chains
introduced in Section \ref{polimeri}. Using a polymer expansion
technique we prove in Theorem \ref{thm:ferro} that when the
parameter $r$ is positive the interaction associated to the
corresponding spin chain model is of ferromagnetic type.

\begin{figure}[h]
\vspace*{-10mm}\begin{center}
\includegraphics[width=10.0cm]{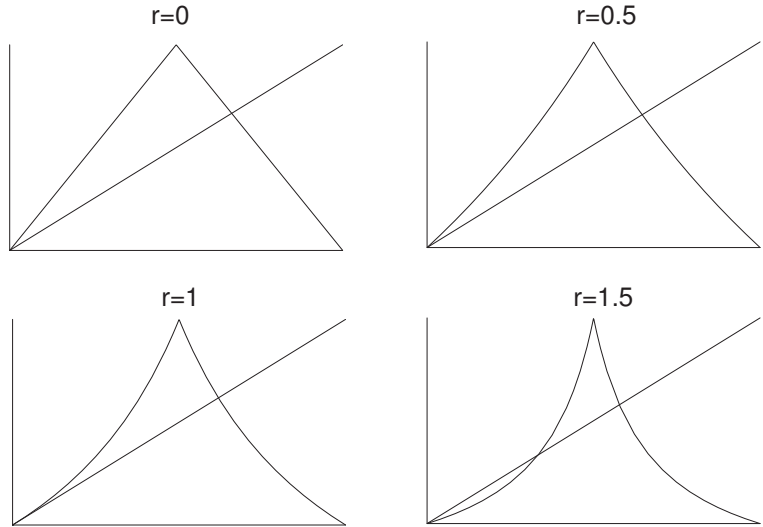}
\caption{{\small The one-dimensional family $F_r$}}
\label{fam}
\end{center}
\end{figure}

\noindent
In the last section we establish
explicit formulas for the iterates of the transfer operators (Propositions \ref{uno}, \ref{caratteri} and \ref{andreaslemma})
which are used to evaluate the canonical (and grand canonical) partition functions as well as some twisted sum with possible
number theoretic significance. From this analysis it turns out that in the canonical setting our models undergo a phase
transition whose features are described in Theorem \ref{fase}.%
 \Section{A one-parameter family of $1D$ maps}\label{family}
\noindent
For
$r\in (-\infty ,2)$, let $F_r$  denote the piecewise real-analytic map
$F_r$ of the interval $[0,1]$ defined as \beqn\label{oneparmap}
F_r(x):= \left\{
\begin{array}{rl} F_{r,0}(x)  &  ,\mbox{ if} \;

 0\leq x\leq 1/2 \\
 F_{r,1}(x)&,  \mbox{ if} \;  1/2 < x \leq 1
 \end{array} \right.
\eeqn
where
$$
F_{r,0}(x)={(2-r)x\over 1-rx}\quad\hbox{and}\quad
F_{r,1}(x)=F_{r,0}(1-x)={(2-r)(1-x)\over 1-r+rx}\cdot
$$
Although some of the result obtained below hold true for a wider
range of $r$ values, in this paper we shall mainly restrict to $r\in
[0,1]$ where this is a Markov family interpolating between the {\sl Farey map}
($r=1$) and the linear expanding {\it tent map} ($r=0$), see Fig.\
\ref{fam}.

\subsection{Previous results}
%
We now recall the main properties of this family,
referring to \cite{GI} for details.
For $r\in[0,2)$ we have\footnote{{\bf Notational warning:} The parameters $r$ and $\rho$, although simply related, are both useful to express the various
quantities introduced in the sequel, and we therefore keep using both of them. Nevertheless, as long as the quantities dealt with below are well defined for all $r <2$
 we shall suppress this parameter.

}
\beq\label{derfix1}
\inf_{x\in [0,1]}|F_r'(x)| = F_{r,0}'(0) =- F'_{r,1}(1)= 2-r=:
\rho.
\end{equation}
This means that for $0\leq r<1$ the map $F_r$ is uniformly expanding, i.e.
$|F_r'|\geq \rho >1$,  thus providing an example of {\sl analytic
Markov map} \cite{Ma}. On the contrary, for $r=1$ one has
$|F'(x)|>1$ for $x>0$ but $F'(0)=1$. For $r>1$ the origin becomes
attractive and there is new repelling fixed point $x^*$ with
$F_r'(x^*)=(2-r)^{-1}$.

\noindent
The left and right inverse branches of $F_r$ are given by
\beq\label{conj}
\Phi_{r,0}:[0,1]\to\left[0,\eh\right] \quad , \quad
\Phi_{r,0}(x)  = {x\over \rho + rx}= {1\over 2}-{1\over 2}\left({\rho-\rho\, x\over \rho +rx}\right)\, ,
\end{equation}
and \beq\label{conj1} \Phi_{r,1}:[0,1]\to\left[{1\over 2},1\right]
\quad , \quad\Phi_{r,1}(x)=1- \Phi_{r,0}(x)= {1\over 2}+{1\over
2}\left({\rho-\rho\, x\over \rho +rx}\right)\cdot
\end{equation}
respectively. The left inverse branch
$\Phi_{r,0}$ is conjugated to the linear map $T_r(x)=\rho \, x+r$ through
the map $J(x)=J^{-1}(x)=1/x$:
\beq
\Phi_{r,0}^n(x) = J^{-1}\circ T_r^n \circ J (x),\quad\forall n\geq 1
\end{equation}
This yields (\cite{GI}, Lemma 2.1):
\beq
\Phi^n_{r,0}(x) =\left(  {\rho^n\over x} + r\,
\sum_{k=0}^{n-1}\rho^k\right)^{-1}.
\end{equation}
\noindent The {\sl Perron-Frobenius operator} or {\sl transfer
operator} $\mathcal{P}_{r}$ associated to $F_r$ is the operator
acting on functions $f:[0,1]\to\C$ as
\begin{eqnarray}
\mathcal{P}_{r} f(x)\,&=&\,\sum_{y: F_r(y)=x}\frac{f(y)}{\vert
F_r'(y)\vert}\\
&=&\,\frac{\rho}{(\rho+rx)^2}\left[ f\left(\frac{x}{\rho +
rx}\right)+f\left(1-\frac{x}{\rho + rx}\right)\right].\nonumber
\end{eqnarray}
\vskip.3cm
\noindent
 For $r\in [0,1]$ the fixed function of $\mathcal{P}_{r}$
corresponds to the density of
 an $F_r$-invariant absolutely continuous measure $\nu_r (dx)$. A
 direct calculation shows that
 \be\label{acim} \nu_r (dx)={K_r\over (1-r+rx)}\, dx\ee with
$K_r$ a suitable positive constant. For $r<1$ the choice
$$
K_r=\left\{\begin{array}{lll} \frac{-r}{\log \,(1-r)}&,&r\in(0,1)\\
1&,&r=0\end{array}\ri.
$$
yields $\nu_r([0,1])=1$ . In particular
$$
\nu_r([0,1/2])=1-\nu_r([1/2,1])=\left\{\begin{array}{lll}
\frac{\log(1-r/2)}{\log(1-r)}&,&r\in(0,1)\\
1/2&,&r=0\end{array}\ri.
$$
We refer to \cite{GI} for several results on the measure $\nu_r$
with different normalization constants.


\noindent Besides the relation between fixed functions and invariant
measures, also the rest of the spectrum ${\rm sp}(\mathcal{P}_{r})$
of the transfer operator is strongly related to the dynamical and
statistical properties of $F_r$ (for a general overview see
\cite{Ba}. On Banach spaces of sufficiently smooth functions, e.g.
$C^k([0,1])$, $\mathcal{P}_{r}$ is {\sl quasi-compact}. Namely, its
spectrum is made out of a finite or countable set of isolated
eigenvalues with finite multiplicity (the discrete spectrum) and the
{\sl essential spectrum} confined in a disk around the origin.

\noindent
More precisely, for the family $F_r$, the following can be proven
\cite{GI}
\begin{proposition}
For all $ r\in [0,1), k\geq 0$, the essential spectrum of
$\mathcal{P}_{r}:C^k\to C^k$ is a disk of radius
$$
r_{\mbox{ess}}(\mathcal{P}_{r})\leq e^{-k\log\rho}.
$$
Moreover, for $r=1$ and for each fixed $k\geq 0$ the essential
spectrum coincides with the whole unit disk.
\end{proposition}
%
\subsection{Generalized transfer operators}
%
\noindent The structure under the "essential spectrum rug" can be
revealed by looking at the action of the operator on Banach spaces
of more regular functions. We shall do it for a more general family
of transfer operators. Given a complex weight $s\in\C$, we let
$\Pqr$ denote the {\sl generalized transfer operator} associated to
the map $F_r$. It acts on a function $f: [0,1] \to \C$ as
$$
\Pqr f (x) := {\rho^s \over (\rho+rx)^{2s}}\left[ f \left({x \over
\rho+rx}\right)+ f \left(1-{x \over \rho+rx}\right) \right].
$$

\noindent It is remarkable that having fixed $s\in\C$ these
operators leave invariant the same Hilbert space for all $r\in
[0,1]$.
\begin{definition}\label{h0}
We denote by ${\cal H}_s$
the Hilbert space of all complex-valued functions $f$ which can be
represented as a generalized Borel transform $$ f(x)=({\B}_s[
\varphi])(x):={1\over x^{2s}}\int_0^\infty  e^{-{t\over x}}\,e^t\,
\varphi (t)\, m_s(dt),\quad \varphi \in L^2(m_s), $$ with inner
product $$(f_1,f_2) := \int_0^\infty  \varphi_1(t)\, {\overline
{\varphi_2  (t)}}\, m_s(dt)\quad\hbox{if}\quad f_i={\B}_s\,
\varphi_i, $$ and measure $$ m_s(dt)= t^{2s-1}\,e^{-t}\, dt .$$
\end{definition}
An alternative representation can be obtained by a simple change of
variable when $x$ is real and positive: $$ x^{2s-1}\cdot f(x) =
\int_0^\infty e^{-t}\, \psi (tx)\, dt\quad \hbox{with}\quad \psi (t)
:= t^{2s-1}\cdot \varphi (t)\, . $$ Note that a function $f \in
{\cal H}_s$ is analytic in the disk
\be\label{disk} D_1
:=
\l\{x\in \C \,|\, \Re\, ({\textstyle {1\over x}}) > \eh\ri\}
=\{x\in \C \,|\, \, |x-1|<1\}. \ee
Partial integration shows that if there is a
sequence $\{a_n\}_{n=0}^\infty$ such that $$ t^{2s-1}\cdot \varphi
(t) =\sum_{n=0}^\infty  {a_n \over n!}\, t^n\quad \hbox{then}\quad
x^{2s-1}\cdot f(x) = \sum_{n=0}^\infty  a_n\, x^n. $$

\noindent To understand the action of the transfer operator on this
Hilbert space, let us define for $r\in[0,1]$ the following two families of operators
$M_{s,r}, N_{s,r}:L^2(m_s)\to L^2(m_s)$ as:
\beq\label{M}
M_{s,r}\, \varphi (t) := {1\over \rho^{s}}\,e^{-{r\over \rho}t}\,
\varphi\left({t\over \rho}\right)
\end{equation}
and
\be\label{N}
N_{s,r}\, \varphi (t) =  {1\over \rho^{s}}\,e^{\left({1-\rho\over
\rho}\right)t}\, \int_0^\infty \frac{J_{2s-1}\left({{2\sqrt{ut}/
\rho}}\right)}{\left(ut/ \rho\right)^{s-1/2}}\, \varphi (u) \,
m_s(du)
\ee
where $J_p$ denotes the Bessel function of order $p$.
Note that for $r=1$ $M_{s,r}$ reduces to a multiplication operator.

\noindent
\begin{proposition}

The space $\mathcal{H}_s$, $s\in\C$  is invariant for ${\Pqr}$ and
we have
\beq\label{Pop}
{\Pqr}\mathcal{B}_s\, [\varphi] = \mathcal{B}_s\,
[(M_{s,r}+N_{s,r})\varphi]\,
\end{equation}
Moreover, for all $r\in [0,1)$ the transfer operator $\Pqr$ when
acting upon $\mathcal{H}_s $  is of the trace-class, with
\beq\label{tracciona}
\mbox{trace}\, (\Pqr) = {\rho^{1-s}\over \rho-1 } +{\rho^s\over \sqrt{1+4\rho}}
\left({2\over 1+\sqrt{1+4\rho}}\right)^{2s-1}
\end{equation}
with $\rho =2-r$.
\end{proposition}
{\bf Proof.} The representation of $\Pqr$ on the space $\mathcal{H}_s$ can be readily obtained by generalizing the calculations done
 in \cite{GI}, Thm 4.3, to prove the statement for $s=1$, by using the {\em Tricomi identity} \cite{GR}
 $$
 {1\over u^{p+1}} \int_0^\infty e^{-{t\over u}} \psi (t) dt =\int_0^\infty e^{-tu}\left( \int_0^\infty \left( {t\over s}\right)^{p\over 2} J_p(2\sqrt{st}) \psi (s) ds \right) dt
 $$
with the choice $p=2s-1$. The formula for the trace will be reobtained and generalized to all iterates of $\Pqr$ with the additional tools
developed later on (see Theorem \ref{trace}). Now it can be derived from the following result
 for the spectrum of the two operators $M_{s,r}$ and $N_{s,r}$ which is a simple generalization of
that given for $s=1$ in \cite{GI}, Propositions 4.5 and 4.6.  \hfill$\Box$

\begin{proposition}

\begin{itemize}
\item
For all $ r\in [0,1)$ and $s\in\C$ the operator
$M_{s,r}:L^2(m_s)\to L^2(m_s)$ is of the {\sl trace class} and its spectrum is
given by
$$
\mbox{\rm sp}(M_{s,r})\,=\,\{\mu_k\}_{k\geq 0}\cup\{0\}\,\mbox{ with }\,
\mu_k=\frac{1}{\rho^{s+k}}.
$$
\item
For all $ r\in [0,1]$ and $s\in\C$ the operator
$M_{s,r}:L^2(m_s)\to L^2(m_s)$ is of the {\sl trace class} and its spectrum is
given by
$$
\mbox{\rm sp}(N_{s,r})\,=\,\{\nu_k\}_{k\geq 0}\cup\{0\}\,\mbox{ with }\,
\nu_k=(-1)^{k} \left({4\rho \over (1+\sqrt{1+4\rho})^2}\right)^{s+k}.
$$
\end{itemize}
 \end{proposition}
\begin{remark}\label{contspec}
For $r=1$ the spectrum of operator $M_{s,1}$ is given by the closure
of the range of the multiplying function $e^{-t}$ that is $[0,1]$.
Therefore, since ${\cal P}_{s,1}$ acts as a selfadjoint compact
perturbation of a selfadjoint multiplication operator, ${\rm
sp}({\cal P}_{s,1})\supseteq [0,1]$ (see \cite{Is}, \cite{Pr}).
\end{remark}

 \noindent We end this section with some remarks on the
general structure of the eigenfunctions of the operator $\Pqr$. The
matrix \be \label{esseerre} S_r:= \bsm
r-1 & 2-r \\
 r & 1-r \\
\esm =
\bsm
  1-\rho & \rho \\
 2-\rho & \rho -1 \\
\esm \in PSL(2,\R)
\ee
with $S^2_r=\mbox{Id}$ and $\mbox{det} \,S_r=-1$ acts on $\C$ as the M\"obius transformation
\be\label{Sr}
 {\hat S}_r(x) :=
{(r-1) x+ 2-r\over rx+1-r}. \ee
Since $\Phi_{r,i}\circ {\hat S}_r =\Phi_{r,1-i}$,
$i=0,1$, we have the implication $$ {\cal P}_{s,r} f =
\lambda \, f, \quad  \lambda
\ne 0 \quad \Longrightarrow  \quad {\cal I}_{s,r} f=f $$
for the involution
$$
({\cal I}_{s,r} f)(x):=\frac{1}{(rx+1-r)^{2s}} f\left({\hat
S}_r(x)\right). $$
Therefore the eigenvalue equation is equivalent to the generalized
three-term functional equation
$$ \rho^{-s} \,\lambda \,f(x)\,=\,
f\left(\frac{x}{\rho}+1\right)+\frac{1}{((2-\rho)x+\rho-1)^{2s}}f
\left(\frac{(\frac{1}{\rho}+{\rho}-1)x+\rho}{(2-\rho)x+\rho-1}\right)
$$
which for $r=1$ reduces to the Lewis-Zagier functional equation arising in the theory of modular forms \cite{LeZa2}.


%
\Section{Dynamical binary trees and coding}\label{alberi}
\subsection{A generalization of the Farey Tree}
%
For each $r\in (-\infty,2)$ we construct a `dynamical' binary tree $\tree(r)$
from the sequences
\beq \label{treen} \tree_{n}(r):=\cup_{k=0}^{n+1}F_r^{-k}(0).
\end{equation} The
ordered elements of $\tree_n$ can be written as ratios of
irreducible polynomials over $\Z$, with positive coefficients when written as functions of the variable
$\rho =2-r$.
For example,
$$\tree_0=\left(\frac{0}{1},\frac{1}{1}\right),\quad \tree_1=
\left(\frac{0}{1},\frac{1}{2},\frac{1}{1}\right),$$
$$\tree_2=\left(\frac{0}{1},\frac{1}{2+\rho},
\frac{1}{2},\frac{1+\rho}{2+\rho},\frac{1}{1}\right),$$
$$
\tree_3\setminus \tree_2=\left(\frac{1}{2+\rho+\rho^2},\frac{1+\rho}{2+3\rho},
\frac{1+2\rho}{2+3\rho},
\frac{1+\rho+\rho^2}{2+\rho+\rho^2}\right)$$
and so on.
\begin{figure}[h]
\begin{center}
\includegraphics[width=6.5cm]{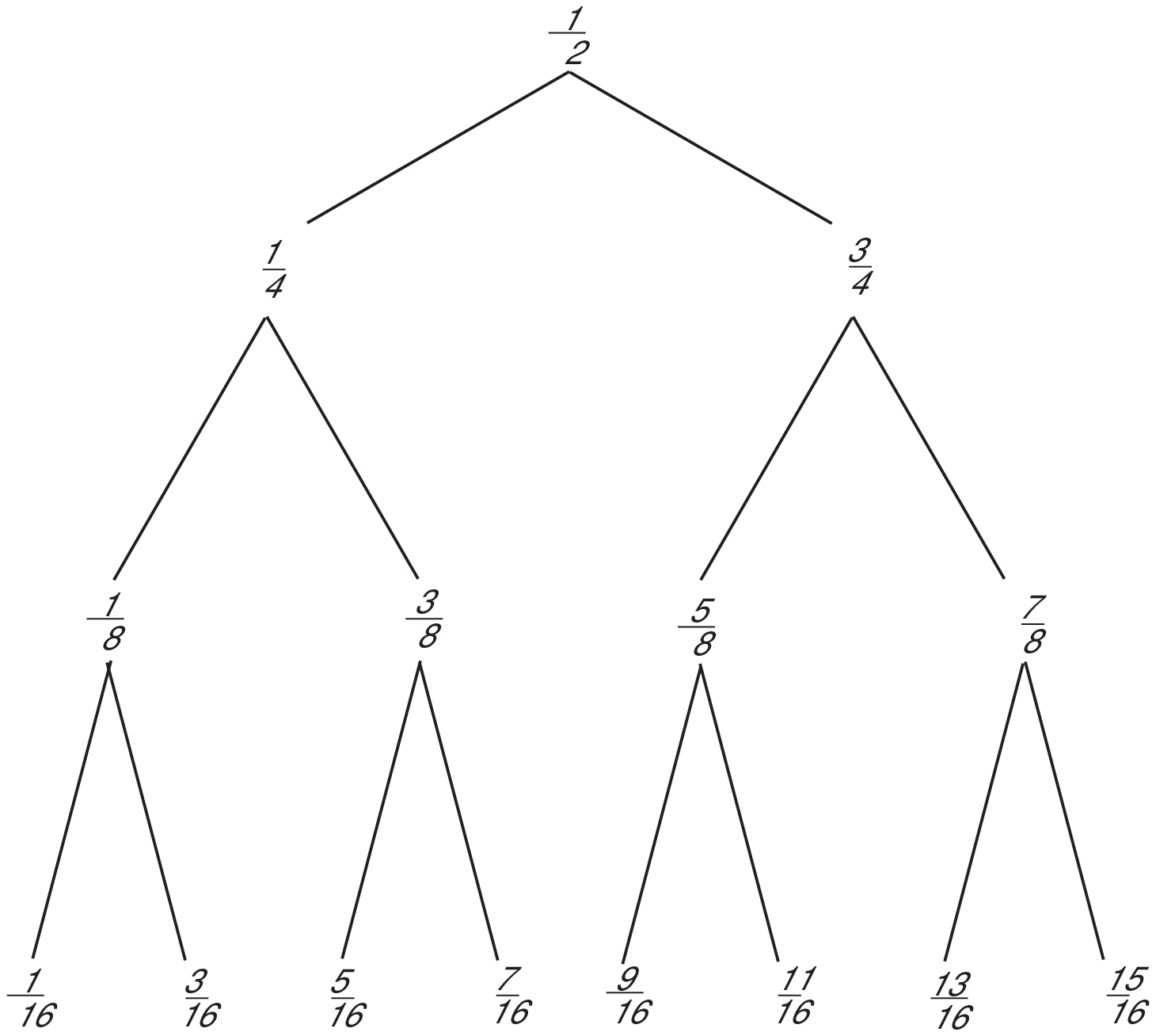}
\includegraphics[width=6.5cm]{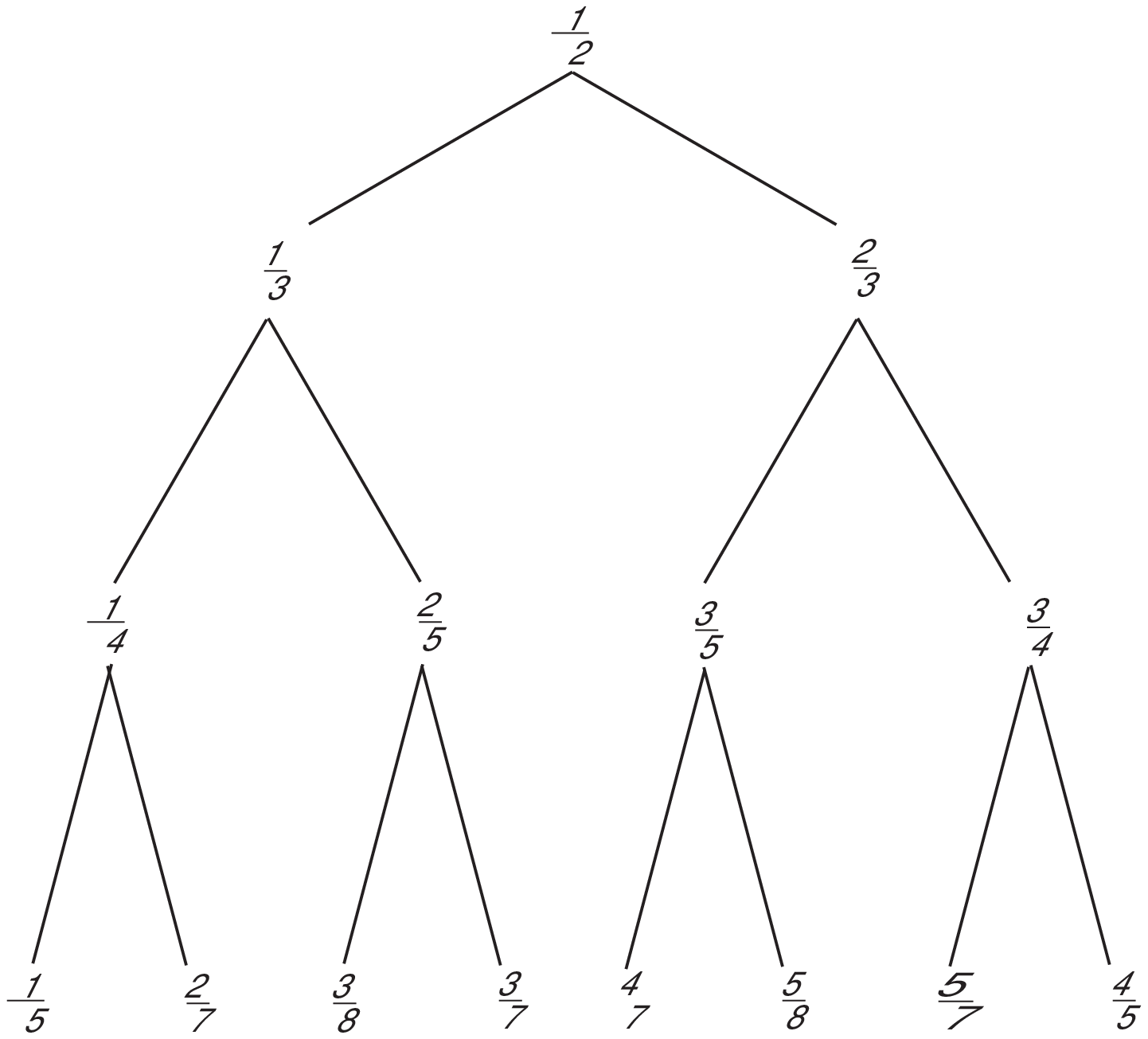}
\caption{{\small The dyadic tree {\it (left)} and Farey tree {\it
(right)}.}}\label{fareytree}
\end{center}
\end{figure}
The rooted tree $\tree $ is now constructed as follows:
\begin{itemize}
\item for $n\geq 1$ the $n$-th row has $2^{n-1}$
vertices, ordered in $[0,1]$, and coincides with
 $\tree_{n}\setminus \tree_{n-1}$.
The vertex $1/2\in\tree_1$ is considered as the root,
\item
edges connect each element in $\tree_n\setminus\tree_{n-1}$ to a pair of elements in
$\tree_{n+1}\setminus\tree_n$ in such a way that no edges overlap. The edge pointing to the left  is labelled by $0$, the other edge by $1$.
 \end{itemize}
We say that an element
${p\over q} \in \tree $ has {\sl rank} $n$, written ${\rm rank}({p\over q})=n$, if it belongs to $\tree_n\setminus \tree_{n-1}$.
We say moreover that two elements ${p\over q}, {p'\over q'}$ in $\tree_n$ are {\sl neighbours} whenever they are neighbours when $\tree_n$ is considered as an ordered subset of $[0,1]$.
It is an easy task to realize that for each pair of neighbours ${p\over q}, {p'\over q'}$ in $\tree_n$ one of them has rank $n$ and the other has rank $n-k$ for some $1\leq k <n$.

\noindent Since $F_r$ is expansive for all $r\in [0,1]$ the vertex
set of $\tree (r)$ is dense in $[0,1]$. In particular (see Fig.\ref{fareytree}),

\begin{itemize}
\item
$\tree(1)$ is
the {\sl Farey tree} whose vertex-set is $\Q \cap (0,1)$.
\item
$\tree(0)$ is the {\it dyadic tree} whose vertex-set is the set of
all dyadic rationals of the form $k/2^m$.
\item
For
$r>1$ the vertex set of $\tree (r)$ is not dense anymore and for
$r\nearrow 2$ accumulates to the single point $1/2$.
\end{itemize}

\begin{figure}[h]
\begin{center}
\includegraphics[width=10.0cm]{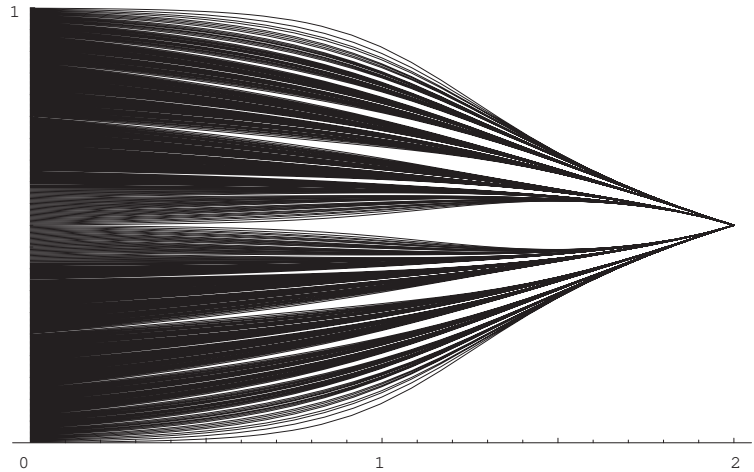}
\caption{{\small Plot of the set ${\cal T}_n\setminus {\cal T}_{n-1}$
for $n=10$ and $0\leq r\leq 2$.}}
\label{fraz}
\end{center}
\end{figure}

In Fig. \ref{fraz} we plot the $n$-th row of $\tree (r)$ for
$n=10$ and different values of $r\in [0,2)$.


\begin{figure}[h]
\begin{center}
\epsfig{file=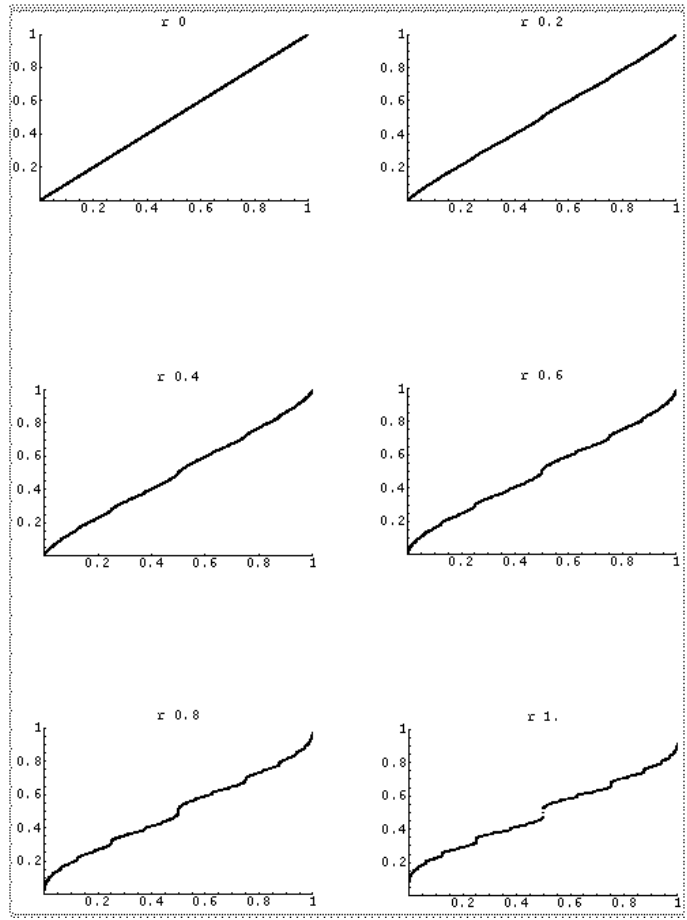,width=9.0cm} \caption{{\small
The conjugating function $h_r^{-1}$ for some values of $r\in [0,1]$.
For $r=1$ we see the graph of the (inverse of the) Minkowski
function $?$.}} \label{conjuga}
\end{center}
\end{figure}

\noindent
The following result is a straightforward consequence of the construction
given above and developed in what follows (see Lemma \ref{codrel}).
\begin{lemma}\label{code} For all $r\in [0,1]$ we have that

\begin{itemize}

\item to every $x\in [0,1]$ there corresponds a unique
sequence $\phi_r (x)\in \{0,1\}^\N$
which represents the sequence of edges of an infinite path $\{x_k\}_{k\geq 1}$ on
$\tree (r)$ with $x_1=1/2$  and $x_k\to x$ (if $x$ is a vertex
of $\tree (r)$ we extend the sequence leading to that vertex either with $01^\infty$ or $10^\infty$);

\item
the map $\phi_r
\circ F_r \circ  \phi_r^{-1}$ acts as the left-shift on
$\Sigma := \{0,1\}^\N/\iota$, where $\iota(\sigma) :={\overline \sigma}$
with ${\overline \sigma_j}:=1-\sigma_j$;

\item
the homeomorphism $h_r:=\phi_0^{-1}\circ \phi_r:[0,1]\to [0,1]$ conjugates $F_r$ to
$F_0$, so that the measure $d h_r(x)$ is $F_r$-invariant and its
entropy is equal to $\log 2$.

\end{itemize}
\end{lemma}
\noindent
The (inverse) conjugating function $h_r^{-1}$ for different values
of $r\in [0,1]$ is sketched in Fig.\ref{conjuga}.

\begin{remark}
$d h_r(x)$ is the measure of maximal entropy for $F_r$, and for
$r\ne 0$ is plainly singular w.r.t.\   Lebesgue (cfr. (\ref{acim})).
\end{remark}
\begin{example}
\begin{itemize}
\item
If for $r=0$ in binary notation $x=0.\sigma=\sum_{n=1}^\infty\sigma_n2^{-n}$ then $\phi_0 (x) =
\sigma$.
\item
Instead, for $r=1$, if in continued fraction notation $x=[a_1,a_2,a_3,\dots]$ then  $\phi_1 (x) =
0^{a_1}1^{a_2}0^{a_3}\cdots$. In this case the conjugating function
$h_1(x)=\phi_0^{-1}\circ \phi_1(x)$ is but the
{\sl Minkowski question mark} function \cite{Mi}, defined as
\begin{eqnarray} ?(x)&:=&\sum_{k\geq 1}(-1)^{k-1}\, 2^{-(a_1+\cdots
+a_k-1)}
 \nonumber \\
&=& 0.{\underbrace
{00\dots 0}_{a_1-1}}\;{\underbrace {11\dots 1}_{a_2}}\;{\underbrace
{00\dots 0}_{a_3}}\; \cdots\nonumber
\end{eqnarray}
\end{itemize}
\end{example}
%


\vskip 0.2cm
\noindent
For $k\in\bN_0$ the point $1/2$ has exactly $2^k$ preimages w.r.t.\   the iterated
map $F_r^k$. We enumerate them using the group ${\bf G}_k:=(\bZ/2\bZ)^k$,
the group elements $\sigma=(\sigma_1,\ldots,\sigma_k)\in{\bf G}_k$ being ordered
lexicographically. Considered as a function of $r$ the preimage indexed by $\sigma$
is a quotient $\frac{p_k(\sigma)}{q_k(\sigma)}$ of polynomials. These are
inductively defined by setting $p_0:=1,\ q_0:=2$,
\beqn\label{A}
p_{k+1}(0,\sigma)&:=&p_k(\sigma),\nonumber\\
p_{k+1}(1,\sigma)&:=&(2-r)q_k(\overline{\sigma})+(r-1)p_k(\overline{\sigma}),
\eeqn
and
\beqn\label{B}
q_{k+1}(0,\sigma)&:=&(2-r)q_k(\sigma)+rp_k(\sigma),\nonumber\\
 q_{k+1}(1,\sigma)&:=&(2-r)q_k(\overline{\sigma})+rp_k(\overline{\sigma}).
\eeqn
Here we use the shortcut $\overline{\sigma}$ for the group element $(1-\sigma_1,\ldots,
1-\sigma_k)$.

\noindent The identities
\beq p_k(\sigma)+p_k(\overline{\sigma}) =
q_k(\sigma) = q_k(\overline{\sigma}),\qquad(\sigma\in{\bf G}_k),
\Leq{simmetria}
follow immediately from (\ref{A}) and (\ref{B}). Note
that in terms of the left and right inverse branches of $F_r$ (cfr
(\ref{conj}) and (\ref{conj1})) we can write
\beq\label{iter}
{p_{k+1}(0,\sigma) \over q_{k+1}(0,\sigma)}=\Phi_{r,0}\left({p_k(\sigma) \over
q_k(\sigma)}\right),\quad
{p_{k+1}(1,\overline{\sigma}) \over q_{k+1}(1,\overline{\sigma})}=\Phi_{r,1}\left({p_k(\sigma) \over
q_k(\sigma)}\right).
\end{equation}
The induction step for the following arithmetic characterization for $r=0$ and $r=1$
is provided by (\ref{iter}).

\begin{lemma}
 Let
$\sigma\in{\bf G}_k$ be of the form
$$
\sigma = ({\underbrace
{0,0,\dots 0,}_{a_1-1}}\;{\underbrace {1,1,\dots 1,}_{a_2}}\;{\underbrace
{0,0,\dots 0,}_{a_3}}\; \cdots \;{\underbrace
{u,\dots ,u}_{a_n-1}\;})
$$
with $u=0$ for $n$  odd and $u=1$ otherwise, and some positive integers $a_i$, $1\leq i\leq n$,
such that $a_n>1$ and
$$
a_1=k+1 \quad \hbox{for}\quad\quad n=1\quad \hbox{and}\quad
\sum_{i=1}^n a_i = k+2 \quad \hbox{for}\quad \quad n>1.
$$
We have
\begin{eqnarray}
r=0\qquad  &\Longrightarrow& \qquad {p_k(\sigma) \over q_k(\sigma)} = 0.\, \sigma \, 1,\nonumber \\
r=1\qquad &\Longrightarrow& \qquad {p_k(\sigma) \over q_k(\sigma)} =
\begin{cases} 1/(a_1+1), &
n=1, \cr
 [a_1,\cdots, a_n], &
n > 1. \cr
\end{cases}\nonumber
\end{eqnarray}
\end{lemma}
\noindent The connection between this coding of the leaves and the
one naturally induced by the dynamics can be understood as follows:
let the group isomorphisms $\Psi_k:{\bf G}_k\to{\bf G}_k$ be given by
$$
\Psi_k(t_1,t_2,\ldots,t_k):=(t_1,t_1+t_2,t_1+t_2+t_3,\ldots,t_1+t_2+\cdots+t_k) \quad
({\rm mod}\ 2).
$$
Clearly
$$
\Psi_k^{-1}(s_1,s_2,\ldots,s_k)=(s_1,s_1+s_2,s_2+s_3,\ldots,s_{k-1}+s_k) \quad
({\rm mod}\ 2).
$$
\begin{lemma}\label{codrel}
For all $r\in (-\infty,2)$ and $\sigma\in{\bf G}_k$,
\beq\label{cod}
\frac{p_k(\sigma)}{q_k(\sigma)}=\Phi_{\Psi_k^{-1}(\sigma)}(\eh),
\end{equation}
with the iterated inverse branch $\Phi_{(t_1,\ldots,t_k)}\,:=\,\Phi_{t_1}\circ
\Phi_{(t_2,\ldots,t_k)}$.
\end{lemma}
{\bf Proof.}
The relation is obviously true for $k=1$ and
$\Psi_1=\mbox{Id}_{{\bf G}_1}$. Assume now that (\ref{cod}) holds
for a given $k\in\N$, then from the relations:
$$
\Psi^{-1}_{k+1}(1,\bar{\sigma})\,=\,(1,\sigma_1,\sigma_1+\sigma_2,\ldots,\sigma_{k-1}+\sigma_k)\,=\,(1,\Psi_k^{-1}(\sigma))
$$
and
$$
\Psi^{-1}_{k+1}(0,{\sigma})\,=\,(0,\sigma_1,\sigma_1+\sigma_2,\ldots,\sigma_{k-1}+\sigma_k)\,=\,(0,\Psi_k^{-1}(\sigma)),
$$
it follows from (\ref{iter}) that
$$
\frac{p_{k+1}(0,\sigma)}{q_{k+1}(0,\sigma)}=\Phi_{0}\circ
\Phi_{\Psi_k^{-1}(\sigma)}(\frac{1}{2})=\Phi_{\Psi_{k+1}^{-1}(0,\sigma)}(\frac{1}{2}),
$$
and
$$
\frac{p_{k+1}(1,\bar{\sigma})}{q_{k+1}(1,\bar{\sigma})}=\Phi_{1}\circ
\Phi_{\Psi_k^{-1}(\sigma)}(\frac{1}{2})=\Phi_{\Psi_{k+1}^{-1}(1,\bar{\sigma})}(\frac{1}{2}).
$$
\hfill $\Box$\\[2mm]
Let now consider again the sequence $\tree_n$ defined in (\ref{treen}). The following result generalizes for $\tree (r)$ the mediant operation which generates the Farey tree $\tree (1)$ \cite{GKP}.
\vskip 0.5cm
\begin{proposition}\label{fareysum} Let $r\in (-\infty,2)$. For each
pair of neighbours ${p\over q} <{p'\over q'}$ in $\tree_n$, with ${\rm
rank}({p\over q})=n-k$ and ${\rm rank}({p'\over q'})=n$, its {\bf child} ${p''\over q''}$ given by
$$
{p''\over q''}:= {p'+\rho^k p \over q'+\rho^k q}
$$
satisfies
$$
{p\over q}<{p''\over q''}<{p'\over q'}
\quad\hbox{and}\quad
{\rm rank}({p''\over q''})=n+1.
$$
Moreover, it holds
$$
p'\, q - p\, q' = \rho ^{n-k},
$$
(the roles of  $p,q, p',q'$ are plainly reversed if ${p'\over q'} < {p\over q}$).
\end{proposition}
{\bf Proof.}
Let us assume that ${p\over q} <{p'\over q'}$ with ${\rm rank}({p\over q})=n-k$ and ${\rm rank}({p'\over q'})=n$. The opposite
case is similar. We have
$$
{p\over q}\equiv {p_{n-k-1}(\sigma)\over q_{n-k-1}(\sigma)}, \quad (\sigma \in {\bf
G}_{n-k-1}),\quad\hbox{and}\quad {p'\over q'}\equiv  {p_{n-1}(\sigma')\over q_{n-1}(\sigma')}, \quad (\sigma' \in {\bf G}_{n-1}).
$$
Moreover, since $p\over q$ and $p'\over q'$ are neighbours we have $\sigma' = \sigma \, \sigma^*$ where $ \sigma^* \in
{\bf G}_{k}$ is given  by $ \sigma^* = (1,0\dots,0)$.

\noindent
Now, the element having rank $n+1$ which appears in $\tree_{n+1}$ between ${p\over q}$ and ${p'\over q'}$ has the form ${p_{n}(\sigma'')\over q_{n}(\sigma'')}$
where $\sigma'' \in  {\bf G}_{n}$ is given by $\sigma''= \sigma \, \sigma^*
0$. Therefore, using Lemma \ref{codrel} we have the expressions
\begin{eqnarray}
{p_{n-k-1}(\sigma)\over q_{n-k-1}(\sigma)} &=& \Phi_{\Psi_{n-k-1}^{-1}(\sigma)}(\frac{1}{2})\nonumber \\
{p_{n-1}(\sigma')\over q_{n-1}(\sigma')} &=& \Phi_{\Psi_{n-k-1}^{-1}(\sigma)}\circ
\Phi_{\Psi_k^{-1}(\sigma^*)}(\frac{1}{2})\nonumber \\ {p_{n}(\sigma'')\over q_{n}(\sigma'')} &=&
\Phi_{\Psi_{n-k-1}^{-1}(\sigma)}
\circ \Phi_{\Psi_k^{-1}(\sigma^*0)}(\frac{1}{2})\nonumber
\end{eqnarray}
A  direct computation yields
$$
\Phi_{\Psi_k^{-1}(\sigma^*)}(\frac{1}{2})={1+2\rho +\sum_{i=2}^{k-1}\rho ^i \over 2+3\rho +2\sum_{i=2}^{k-1}\rho ^i }=:{a'\over
b'}
$$
and, setting ${a\over b}:={1\over 2}$, we get
$$
\Phi_{\Psi_k^{-1}(\sigma^*0)}(\frac{1}{2})={1+2\rho +\sum_{i=2}^{k}\rho ^i \over 2+3\rho +2\sum_{i=2}^{k}\rho ^i }=:{a''\over b''}=
{a'+\rho ^k a \over b'+  \rho^k b}.
$$
Therefore, to complete the proof it suffices to verify that the
three ratios ${a\over b}, {a'\over b'}, {a''\over b''}$ verify
$$
{a'' \over b''}= {a'+\rho ^k a \over b'+\rho^k b}
$$
if and only if, for any choice of $(t_1,\ldots,t_l)\in {\bf G}_l$,  the ratios
$${p\over q}:=\Phi_{(t_1,\ldots,t_l)}({a\over b}),\quad {p'\over q'}:=\Phi_{(t_1,\ldots,t_l)}({a'\over b'}), \quad {p''\over q''}:=\Phi_{(t_1,\ldots,t_l)}({a''\over b''})$$
satisfy
$$
{p'' \over q''}= {p'+\rho ^k p \over q'+\rho^k q}
$$
as well. On the other hand, this property can be easily verified by induction using the expressions
$$
\Phi_0({a\over b}) = {a \over 2a +\rho (b-a)}\quad\hbox{and}\quad \Phi_1({a\over b}) = {a+\rho (b-a) \over
 2a +\rho (b-a)}.
$$
The first statement now follows by taking $l=n-k-1$ and $(t_1,\ldots,t_l)=\Psi_{n-k-1}^{-1}(\sigma)$. The second
follows in a similar way. \hfill $\Box$\\[2mm]

\noindent
Now, for $r\in[0,2)$ the matrices
\beq
  I_j:=\left(
  \begin{array}{cc}
  -j\rho & j\rho\\
  -\rho & \rho
  \end{array}\right),\qquad j=0,1,
\end{equation}
are in ${\rm GL}(n,\R)$. As such $\R$ they act on $\C$ as
M\"obius transformations, and the action of two operators forming the transfer operator ${\cal P}_{s,r}={\cal
P}_{s,r}^{(0)}+{\cal P}_{s,r}^{(1)}$ of
$F_r$ on a function $f:[0,1]\to \C$ can be written as
\beq\label{op1}
({\cal P}_{s,r}^{(j)} f)(x) := \left| \Phi'_j(x) \right|^{s}
f\left(\Phi_j(x)\right),\qquad j=0,1,
\end{equation}
with
\beq\label{op2}
\Phi_j(x):={\hat I}_j(x)={(1-j\rho)x+j\rho \over (2-\rho)x+\rho} \cdot
\end{equation}
Using the involution $S=S_r$ introduced in (\ref{esseerre}) we shall see that the elements of $\tree (r)$  can be
represented by means of a subgroup of ${\rm GL}(2,\R)$ with generators
\beq\label{op3}
L:=I_0=\left(
  \begin{array}{cc}
   1 & 0\\
  2-\rho & \rho
  \end{array}\right) \quad\hbox{and}\quad
  R:=S L S=\left(
\begin{array}{cc}
  1 & \rho\\
  0 & \rho
  \end{array}\right)\cdot
\end{equation}
Note that
\beq\label{commuta}
I_1=S\, R=L\, S.
\end{equation}
For example we have
$$
LL = \left(
  \begin{array}{cc}
  1 & 0\\
  2+\rho-\rho^2 & \rho^2
  \end{array}\right) \quad\hbox{and}\quad
LR=\left(
\begin{array}{cc}
  1 & \rho\\
  2-\rho & 2\rho
  \end{array}\right),
$$
so that
$$
{\hat L}(1)={1\over 2},\quad {\widehat {LL}}(1)={1\over 2+\rho},\quad
{\widehat {LR}}(1)={1+\rho \over 2+\rho}.
$$
More generally, we have the following characterization of the leaves of $\tree (r)$ as matrix products
which generalize what is known for the arithmetic case $r=1$
\cite{Kn1}.
\begin{proposition}\label{codrel1}
For all $r\in (-\infty,2)$ the element
$p_k(\sigma)/q_k(\sigma)$ of $\tree (r)$
can be uniquely presented as
the product $X=L\prod_{i=1}^k M_i$
where $M_i=(1-\sigma_i)L+\sigma_i R$. More precisely,
$$
\frac{p_k(\sigma)}{q_k(\sigma)}= {\hat X}(1).
$$
Since ${\rm det}\, L ={\rm det}\, R =\rho$, we have
${\rm det}\, X = \rho ^{k+1}$.
\end{proposition}
{\bf Proof.}
By Lemma \ref{codrel}, (\ref{op2}), (\ref{op3}) and
(\ref{commuta}) the element $p_k(\sigma)/q_k(\sigma)$  can be written as
\beqno
\lefteqn{\frac{p_k(\sigma)}{q_k(\sigma)}=\Phi_{\Psi_k^{-1}(\sigma)}\left(\eh\right)=
\hat{I}_{\sigma_1}\circ\hat{I}_{\sigma_1+\sigma_2}\circ\ldots\circ\hat{I}_{\sigma_{k-1}+\sigma_k}
\left(\eh\right)}\\
&=&(\hat{L}\circ \hat{S}^{\sigma_1})\circ(\hat{L}\circ\hat{S}^{\sigma_1}\circ S^{\sigma_2})\circ
\ldots\circ(\hat{l}\circ\hat{S}^{\sigma_{k-1}}\circ\hat{S}^{\sigma_k})\left(\eh\right)\\
&=& \hat{L}\circ\hat{M}_{\sigma_1}\circ\ldots\circ\hat{M}_{\sigma_{k-1}}\hat{S}^{\sigma_k}\left(
\eh\right)=\hat{X}\left(\eh\right),
\eeqno
since $\hat{S}^{\sigma_k}\left(\eh\right)=\hat{S}^{\sigma_k}\hat{I}_{\ov{\sigma}_k}(1)=
\hat{M}_{\sigma_k}(1)$.
\hfill $\Box$
%
\subsection{Traces and dynamical partition functions}
%
Let us now define two maps $T_j: \tree (r) \to \R$, $j=0,1$, as
\beq\label{tracce}
T_0\left({p\over q}\right) := \mbox{ trace}(X),\qquad {T_1}\left({p\over q}\right) := \mbox{
trace}(XS)
\end{equation}
where ${p\over q}= {\hat X}(1)$ is the presentation from Prop.\
\ref{codrel1}. For $r=1$ the numbers $T_0\left({p\over q}\right)$ and $T_1\left({p\over
q}\right)$ are given by $p'+q''$ and $p''+q'$, respectively, with ${p\over q}={p'+p''\over q'+q''}$.

\noindent
Also note that if ${p\over q}$ occurs in $\tree (r)$ at level $n$, namely ${p\over q}\in \tree_n\setminus
\tree_{n-1}$, then
\beq\label{dete}
\mbox{det}(X)=-
\mbox{det}(XS)=\rho^n.
\end{equation}
\begin{lemma}
$
{T_0}\left({p\over q}\right)+T_1\left({p\over q}\right)=rp+\rho q.
$
\end{lemma}
{\bf Proof.} $T_0\left({p\over q}\right)+T_1\left({p\over q}\right)=
\tr(X(\idty+S)) =\tr\left( \bsm p'&p''\\ q'& q''\esm \bsm
r&\rho\\r&\rho\esm\right)=rp+\rho q$. \hfill $\Box$
\\[2mm]
We already know that the operator ${\cal P}_{s,r}={\cal P}_{s,r}^{(0)}+{\cal P}_{s,r}^{(1)}$ when acting on
the Hilbert space ${\cal H}_s$  is of the trace class for all $r\in [0,1)$ and $s\in \C$. The above
construction provides a closed expression for the trace of ${\cal P}_{s,r}^n:{\cal H}_s \to {\cal H}_s$,
$n\geq 1$.
\begin{theorem} \label{trace} For all $r\in [0,1)$, $s\in \C$ and $n\geq 1$ we have
\beqno
\lefteqn{\hspace*{-8mm}{\rm trace}\,({\cal P}_{s,r}^n)=}\\
&&\hspace*{-18mm}\sum_{{p\over q}\in \tree_n\setminus \tree_{n-1}}\sum_{j=0,1} \,
\frac{\rho^{\, n s}}{\sqrt{T_j^2({p\over q})-(-1)^j4\rho^n}}
\left(
\frac{2}{T_j({p\over q})+\sqrt{T_j^2({p\over q})-(-1)^j4\rho^n}} \right)^{2s -1}
\eeqno
\end{theorem}
{\bf Proof.} We have $2^n$ terms
$$
{\rm trace}\,({\cal P}_{s,r}^n)= \sum_{\sigma \in {\bf G}_n} {\rm trace}\, ({\cal P}_{s,r}^{(\sigma)})
\quad\hbox{with}
\quad
{\cal P}_{s,r}^{(\sigma_1,\dots, \sigma_n)}:={\cal P}_{s,r}^{(\sigma_1)}\cdots {\cal P}_{s,r}^{(\sigma_n)},
$$
and to each of them we can associate the matrix product $I_{\sigma_1}\cdots I_{\sigma_n}$ according to
(\ref{op1}) and (\ref{op2}). On the other hand, the commutation rules (\ref{commuta}) yield
$$
{\underbrace {I_1 I_0\dots I_0}_{k}}\;I_1 = LS \,{\underbrace
{L\dots L}_{k-1}}\;SR = L \,{\underbrace
{R\dots R}_{k}}.
$$
Using this fact it is not
difficult to realize that each term where $I_1$ appears an even number of times can be expressed in the form
$X=L\prod_i M_i$ and, moreover, to each such term there corresponds exactly another term where the number of
occurrences of $I_1$ is odd and which can be written as
$XS$.

\noindent
Finally, for all $r\in [0,1)$ and $s\in \C$ the generic term ${\cal P}_{s,r}^{(\sigma)}$ is a composition
operator of the form
$f(x)
\to |\psi'(x)|^s f(\psi(x))$ where $\psi(x)=\Phi_{(\sigma)}(x)$ is holomorphic and strictly contractive in
a disk containing the unit interval,
with a unique fixed point
$\bar x \in [0,1]$. Standard arguments then yield for
its trace the expression
$|\psi'(\bar x)|^s/(1-\psi'(\bar x))$  (see, e.g., \cite{Ma}, Sect.\
7.2.2). The thesis now follows by direct computation
putting together the above along with (\ref{tracce}) and (\ref{dete}). \hfill $\Box$\\[2mm]

\vskip 0.2cm
\noindent
We now define the signed operator ${\tilde {\cal P}}_{s,r}:{\cal H}_s \to {\cal H}_s$ as
\be
{\tilde {\cal P}}_{s,r}={\cal P}_{s,r}^{(0)}-{\cal P}_{s,r}^{(1)}.
\ee
An immediate consequence of the proof of Theorem \ref{trace} is the following
\begin{corollary}
For all $r\in [0,1)$, $s\in \C$ and $n\geq 1$ we have
\beqno
\lefteqn{\hspace*{-8mm}
{\rm trace}\,({\tilde {\cal P}}_{s,r}^n)=}\\
&&\hspace*{-18mm}\sum_{{p\over q}\in \tree_n\setminus \tree_{n-1}} \sum_{j=0,1} \,
\frac{(-1)^j\rho^{\, n s}}{\sqrt{T_j^2({p\over q})-(-1)^j4\rho^n}}
\left(
\frac{2}{T_j({p\over q})+\sqrt{T_j^2({p\over q})-(-1)^j4\rho^n}} \right)^{2s -1}
\eeqno
so that
\beqno
\lefteqn{\hspace*{-8mm}
{\rm trace}\,({\cal P}_{s,r}^n)+{\rm trace}\,({\tilde {\cal
P}}_{s,r}^n)=}\\
&&\hspace*{-8mm}2\sum_{{p\over q}\in \tree_n\setminus \tree_{n-1}}
\frac{\rho^{\, n s}}{\sqrt{T_0^2({p\over q})-4\rho^n}}
\left(
\frac{2}{T_0({p\over q})+\sqrt{T_0^2({p\over q})-4\rho^n}} \right)^{2s -1}\cdot
\eeqno
\end{corollary}
Furthermore, let us define a dynamical partition function $\Xi_n(s)$
as \be \Xi_n(s):= \sum_{x=F_r^n(x)} \left| (F_r^n)'(x)\right|^{-s}.
\ee Another simple consequence of the above is the following
\begin{corollary}\label{relatio}
\begin{eqnarray}
\Xi_n(s) &=& {\rm trace}\,({\cal P}_{s,r}^n)-{\rm trace}\,
({\tilde {\cal P}}_{s+1,r}^n) \label{prima}\nonumber \\
&=&\sum_{{p\over q}\in \tree_n\setminus \tree_{n-1}}\sum_{j=0,1} \,
{4^s\rho^{\, n s} \over \left( T_j({p\over q})+\sqrt{T_j^2({p\over q})-(-1)^j4\rho^n}
\right)^{2s}}\,\cdot \nonumber
\end{eqnarray}
\end{corollary}
{\bf Proof.}
For $\sigma \in {\bf G}_n$ set $|\sigma|=\sum_{i=1}^n \sigma_i$. The trace of the operator
$(-1)^{|\sigma|}{\cal P}_{s,r}^{(\sigma)}$ has the expression
$(-1)^{|\sigma|}|\psi'({\overline x})|^s/(1-\psi'(\bar x))$ where  $\psi(x)=\Phi_{(\sigma)}(x)$ and
${\overline x}$ is the unique solution of $\Phi_{(\sigma)}(x)=x$ in $[0,1]$. The first identity now follows
from the equality
$$
{|\psi'({\overline x})|^s\over 1-\psi'(\bar x)} - {(-1)^{|\sigma|}|\psi'({\overline x})|^{s+1}\over
1-\psi'(\bar x)}=|\psi'({\overline x})|^s,
$$
and the second by direct calculation. \hfill $\Box$

\begin{remark}
If $X=L^n$ then
\[T_0=1+\rho^n\qmbox{and}T_1=1+\rho +\rho^2+\cdots \rho^{n-1}.\]
Therefore, as $r\nearrow 1$ we have $T_0\to 2$ and $T_1\to n^2+4$.
In particular $\sqrt{T_0^2-4\rho^n}=\rho^n-1 \to 0$ and the
corresponding term in the trace of  ${\cal P}_{s,r}^n$ diverges (see
Remark \ref{contspec}). On the other hand, one easily sees that for
each $n\geq 1$ this is the only term which diverges as $r\nearrow
1$. Unlike traces, the function $\Xi_n(s)$ is well defined for
$r=1$.
\end{remark}
One can store the numbers ${\rm trace}\,({\cal P}_{s,r}^n)$ and
$\Xi_n(s)$ to form the Fredholm determinant \be {\rm det}(1-z\,{\cal
P}_{s,r}):=\exp \left(-\sum_{n\geq 1}{z^n\over n}{\rm trace}\,({\cal
P}_{s,r}^n) \right) \ee and the dynamical zeta function \be \zeta_r
(z,s):= \exp \left(\sum_{n\geq 1}{z^n\over n}\, \Xi_n(s)\right) \ee
respectively. Another consequence of the above  is the following
\begin{corollary} For $r\in [0,1)$ and
$s\in \C$ \be \zeta_r (z,s)={  {\rm det}(1-z\,{\tilde {\cal
P}}_{s+1,r})\over {\rm det}(1-z\,{\cal P}_{s,r})} \cdot \ee
Moreover
the above determinants are entire functions of both $s$ and $z$ and
therefore, for all $s\in \C$, $\zeta_r (z,s)$ is meromorphic in the
whole complex $z$-plane and analytic in $\{z\in \C : z^{-1}\notin
{\rm sp}({\cal P}_{s,r})\}$.
\end{corollary}

 \Section{Polymer model analysis of the Markov family}\label{polimeri}
%
 \noindent The Fourier transform of a function $f:{\bf
G}_k\to\bC$ is
\[\hat{f}:{\bf G}_k\to\bC \qmbox{,} \hat{f}(t):=2^{-k}\sum_{\sigma\in{\bf G}_k}
f(\sigma)(-1)^{\sigma\cdot t}.\]
We now calculate the polynomials $\hat{p}_k(t),\ \hat{q}_k(t)$ for $t\in{\bf G}_k$,
using the language of polymer models.
As in \cite{GuK} we call the group elements $t\in{\bf G}_k$ a {\em polymer} if
$t=(t_1,\ldots,t_k)$ contains exactly one or two ones. So the set of polymers in
${\bf G}_k$ is $P_k:=P_k^{e}\cup P_k^{o}$
with {\em even} resp.\ {\em odd} polymers
\beqn\label{pol}
P_k^{e}&:=&\{p_{\ell,r}:=\delta_\ell+\delta_r\in{\bf G}_k : 1\leq\ell<r\leq k\},\nonumber \\
P_k^{o}&:=&\{p_\ell:=\delta_\ell\in{\bf G}_k : 1\leq\ell\leq k\}.
\eeqn
Slightly diverging from \cite{GuK}, we defined their {\em supports} by
\[\supp(p_{\ell,r}):=\{i : \ell\leq i\leq r\} \qmbox{and} \supp(p_\ell):=
\{1,\ldots,\ell\}.\]
Two polymers are called {\em incompatible} if their supports have nontrivial
intersection. Thus we can uniquely decompose every group element $t\in{\bf G}_k$
as a sum $t=\gamma_1+\ldots+\gamma_n$ of mutually incompatible polymers $\gamma_i
\in P_k$.
The {\em activities} of the polymers are defined as the rational functions
\beq
\hspace*{-10mm}z(p_{\ell,r}):=-\frac{r}{2-r}\l(\frac{2-r}{4-r}\ri)^{\supp(p_{\ell,r})}
\qmbox{,}
z(p_\ell):=-\l(\frac{2-r}{4-r}\ri)^{\supp(p_\ell)}.
\Leq{activity}
\begin{proposition}\label{prop:7.2}
For the polymer decomposition $t=\gamma_1+\ldots+\gamma_{n(t)}$ of
$t\in{\bf G}_k$ the Fourier coefficients equal \beq
\hp_k(t)=\l(\frac{4-r}{2}\ri)^k\prod_{i=1}^{n(t)}z(\gamma_i) \Leq{C}
\beq
\hq_k(t)=\l(\frac{4-r}{2}\ri)^k(1+(-1)^{|t|})\prod_{i=1}^{n(t)}z(\gamma_i)
\Leq{D}
\end{proposition}
{\bf Proof.}
For $k=0$ the above formulae reduce to $\hp_0=p_0=1,\ \hq_0=q_0=2$. The induction
step from $k$ to $k+1$ proceeds by decomposing the group elements in the form
$(\tau,t)\in{\bf G}_1\times{\bf G}_k\cong{\bf G}_{k+1}$.
\\[2mm]
Then by (\ref{A}) and (\ref{B})
\beqno
\lefteqn{\hspace*{-10mm}\hp_{k+1}(\tau,t)=
\eh \Big(\hp_k(t)+(-1)^{\tau+|t|}((2-r)\hq_k(t)+(r-1)\hp_k(t)) \Big)}\\
&\hspace*{-20mm}=&\hspace*{-10mm}\eh\l(\frac{4-r}{2}\ri)^k\prod_{i=1}^{n(t)}z(\gamma_i)
 \Big(1+(-1)^{\tau+|t|} \l((1+(-1)^{|t|})(2-r)+(r-1) \ri) \Big)\\
&\hspace*{-20mm}=&\hspace*{-10mm}\eh \l(\frac{4-r}{2}\ri)^k\prod_{i=1}^{n(t)}z(\gamma_i)\cdot\left\{\begin{array}{cc}
r+\frac{1+(-1)^{|t|}}{2}(4-2r)\,,&\,\tau+|t|=0\ ({\rm mod}\ 2)\\
2-r-\frac{1+(-1)^{|t|}}{2}(4-2r)\,,&\,\tau+|t|=1\
({\rm mod}\ 2)\end{array}\ri.. \eeqno For all $(\tau,t)$ this coincides with
formula (\ref{C}).
\\[2mm]
Similarly we get
\beqno
\lefteqn{\hspace*{-10mm}\hq_k(\tau,t)=\eh \l(1+(-1)^{\tau+|t|}\ri)[(2-r)\hq_k(t)+r\hp_k(t)]}\\
&\hspace*{-20mm}=&\hspace*{-10mm}\eh \l(1+(-1)^{\tau+|t|}\ri)\l(\frac{4-r}{2}\ri)^k\prod_{i=1}^{n(t)}z(\gamma_i)
\l[r+\frac{1+(-1)^{|t|}}{2}(4-2r)\ri],
\eeqno
coinciding with (\ref{D}).
\hfill $\Box$\\[2mm]
This formula for the Fourier coefficients implies a symmetry of the
denominator function $q_k$
which, unlike (\ref{simmetria}) is not immediate from its definition.

\begin{corollary}
For all $r\in [0,2)$, $k\in\bN$ and
$\sigma=(\sigma_1,\ldots,\sigma_k)\in\Gb_k$,
\[q_k(\sigma_k,\sigma_{k-1},\ldots,\sigma_2,\sigma_1) =
q_k(\sigma_1,\sigma_2,\ldots,\sigma_{k-1},\sigma_k).\]
\end{corollary}
{\bf Proof.}
This statement is equivalent to the one
\[\hq_k(t_k,t_{k-1},\ldots,t_2,t_1) =
\hq_k(t_1,t_2,\ldots,t_{k-1},t_k)\qquad(t\in\Gb_k)\] for the Fourier
coefficients. Since anyhow $\hq_k(t)=0$ for odd $|t|$, in formula
(\ref{D}) for $\hq_k$ only activities $z(p_{l,r})$ of even polymers
$p_{l,r}$ appear. Unlike the odd polymers, they have the symmetry
$z(p_{k-r+1,k-l+1}) = z(p_{l,r})$,
which leads to the above statement. \hfill $\Box$\\[2mm]

\vskip 0.2cm \noindent
\subsection{A 1D spin chain model}

In the same spirit as \cite{Kn2} and \cite{Kn3} we now interpret the
sequences $\sigma \in {\bf G}_k$ as different configurations of a
chain of $k$ classical binary spins with energy function
\be\label{energia} Q_k:=\log q_k \, :{\bf G}_k\to\bR. \ee The
corresponding (canonical) partition function will be \be
\label{partfct} {\Zc}(s):=1+\sum_{0\leq k< n}\sum_{\sigma \in {\bf
G}_k}\exp{(-s\, Q_k(\sigma))}\equiv \sum_{{p\over q}\in
\tree_{n}(r)\setminus \{0\}}q^{-s} . \ee

\begin{figure}[h]
\begin{center}
\includegraphics[width=8cm]{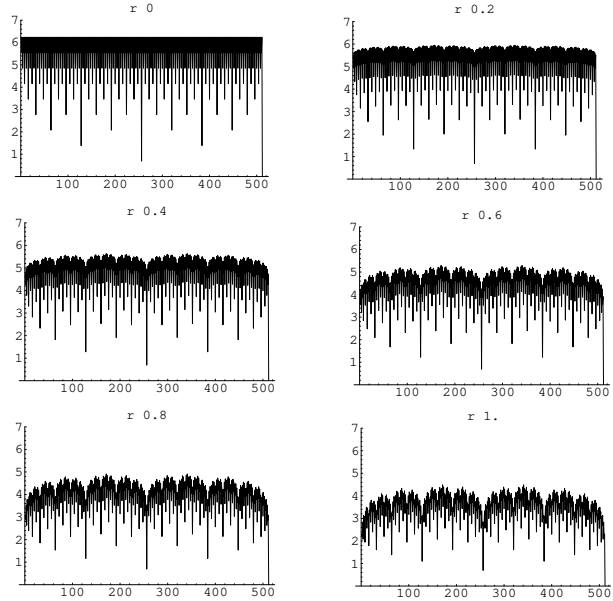}
\caption{{\small The energies $Q_k=\log q_k$ for $k=10$ and
different values of $r\in [0,1]$.}} \label{energies}
\end{center}
\end{figure}

\noindent
Plots of the function $Q_k$ for different values of $r$ are reported in
Fig.\ref{energies}.
\noindent

\noindent
 The canonical partition function can be expressed in a
more standard way using the following

\begin{defi}\label{pqc}
For all $k\in\bN$ we inductively define polynomials
$\pc_k(\si),\qc_k(\si)\in\bZ
[\rho]\ (\si\in\Gk)$ by setting
\[\pc_1(0):=0 \qmbox{,} \pc_1(1):=1 \qmbox{,} \qc_1(0):=1 \qmbox{,} \qc_1(1):=2
\qmbox{,}\] and for $\si\in\Gk$ with
\[r:\Gk\to\bN \qmbox{,}r(\si):=\l\{\begin{array}{cll}
\max\{i|\si_i=1\}&,&\si\in\Gk\backslash\{0\}\\
0&,&\si=0\end{array}\ri.\]
\[\qc_{k+1}(\si,\tau):=\l\{\begin{array}{lll}
\qc_k(\si)&,&\tau=0\\
\rho^{k-r(\si)}\qc_k(\si)+\rho^{k-r(\ov{\si})}\qc_k(\ov{\si})&,&\tau=1
\end{array}\ri.\]
\[\pc_{k+1}(\si,\tau):=\l\{\begin{array}{lll}
\pc_k(\si)&,&\tau=0\\
\rho^{k-r(\si)}\qc_k(\si)+\rho^{k-r(\ov{\si})}(\qc_k(\ov{\si})-
\pc_k(\ov{\si}))&,&\tau=1
\end{array}\ri.\]
\end{defi}
\begin{example}
$\pc_2(0,1)=1,\ \pc_2(1,1)=1+\rho,\ \qc_2(0,1)=\qc_2(1,1)=2+\rho$.
\end{example}

\noindent
The relations between these new polynomials and the old ones (see
(\ref{A})-(\ref{B})) are given by

\begin{equation}\label{nuovipoli}
p_k(\si)=\pc_{k+1} (\si,1) \qmbox{and} q_k(\si)=\qc_{k+1} (\si,1),
\end{equation}
which can be seen to be another way to state Proposition
\ref{fareysum}.

\noindent
Finally, using the new denominators just introduced, we can write
\be\label{cano2} \Zc(s)\,=\, \sum_{\sigma\in{\bf G}_n}
\left(q_n^c(\sigma)\right)^{-s}. \ee

\vskip 0.2cm \noindent
\subsection{Positivity of the interaction and exponential sums}
The negative Fourier coefficients
$-\hat{Q}_k(t)$ can then be interpreted as interaction coefficients
in the sense of statistical mechanics.
\begin{theorem}\label{thm:ferro}
The interaction is ferromagnetic for $r\in[0,1]$, that is
\[-\hat{Q}_k(t)\geq 0, \qquad (t\in{\bf G}_k\backslash\{0\}).\]
\end{theorem}
{\bf Proof.}
We introduce the notation of polymer models
(see, e.g., Gallavotti, Martin-L\"{o}f and Miracle-Sol\'{e} \cite{GMM},
Simon \cite{Si} and Glimm and Jaffe \cite{GJ}).

\noindent
In an abstract setting one starts with a finite set $P$
of {\em polymers}.
Two given polymers $\gamma_1,\gamma_2\in P$ may or may not
be {\em incompatible}. Incompatibility is assumed to be a
reflexive and symmetric relation on $P$.

\noindent Thus one may associate to a $n$--{\em polymer}
$X:=(\gamma_1,\ldots,\gamma_n)\in P^n$ an undirected graph
$G(X)=(V(X),E(X))$ without loops with vertex set
$V(X):=\{1,\ldots,n\}$, vertices $i\neq j$ being connected by the
edge $\{\gamma_i,\gamma_j\}\in E(X)$ if $\gamma_i$ and $\gamma_j$
are incompatible. Accordingly the $n$--polymer $X$ is called {\em
connected} if $G(X)$ is path-connected and {\em disconnected} if it
has no edges ($E(X)=\emptyset$).

\noindent
The corresponding subsets of $P^n$ are called $C^n$ resp.\ $D^n$,
with $D^0:=P^0:=\{\emptyset\}$ consisting of a single element. Moreover
$P^{\infty}:= \bigcup_{n=0}^{\infty} P^n$ with the subsets
$D^{\infty}:= \bigcup_{n=0}^{\infty} D^n$ and
$C^{\infty}:= \bigcup_{n=1}^{\infty} C^n$.
We write $|X|:=n$ if $X\in P^n$.

\noindent
Statistical weights or activities $z:P\ar\bC$ of the
polymers are multiplied to give the activities
$z^X:=\prod_{i=1}^{|X|} z(\gamma_i)$ of multi-polymers.

\noindent
A system of statistical mechanics is called {\em polymer model} if
its partition function $Z $ has the form
\beq
Z  = \sum_{X\in D^{\infty} } \frac{z^X}{|X|!}.
\label{eq:parti}
\end{equation}
Then the free energy is given by
\beq
\log(Z ) = \sum_{X\in C^{\infty} } \frac{n(X)}{|X|!}z^X,
\label{eq:freie}
\end{equation}
with $n(X):=n_+(X)-n_-(X)$, $n_\pm (X)$ being the number of
subgraphs of $G(X)$ connecting the vertices of $G(X)$ with an even
resp.\ odd number of edges.

\noindent
In the present context we index by the number $k$ of spins
and use the polymer set $\cP_k$ from (\ref{pol}).
Then the map
\[D_k^{\infty}\ar {\bf G}_k\qmbox{,}
(\gamma_1,\ldots,\gamma_n)\mapsto \sum_{i=1}^n \gamma_i\]
is a set-theoretic isomorphism between the disconnected multi-polymers
in $\{1,\ldots,k\}$ and the abelian group ${\bf G}_k$.
Similarly (using a superscript $e$ for objects derived from the subset
$P_k^e\subset P_k$ of even polymers)  the image of
\[D_k^{\infty,e}\ar {\bf G}_k\qmbox{,}
(\gamma_1,\ldots,\gamma_n)\mapsto \sum_{i=1}^n \gamma_i\]
is the subgroup of ${\bf G}_k$ whose elements have an even number of ones
\\[2mm]
In terms of this notation and with (\ref{D})
\beqno
\lefteqn{\hspace*{-10mm}\hat{Q}_k(t) =2^{-k}\sum_{\sigma\in{\bf G}_k}\log(q_k(\sigma))(-1)^
{\sigma\cdot t}}\\
&\hspace*{-20mm}=&\hspace*{-10mm} 2^{-k}\sum_{\sigma\in{\bf
G}_k}\log\l(\sum_{s\in{\bf G}_k}\hq_k(s)(-1)^
{s\cdot\sigma}\ri)\cdot(-1)^{\sigma\cdot t}\\
&\hspace*{-20mm}=&\hspace*{-10mm}
\delta_{t,0}\cdot \l(\log(2) +  k\log\l(\frac{4-r}{2}\ri)\ri) +
2^{-k} \sum_{\sigma\in {\bf G}_k} \log\l[\sum_{X\in D_k^{\infty,e}}
\frac{\tilde{z}_\sigma^X}{|X|!}\ri]
\cdot(-1)^{\sigma\cdot t}
\eeqno
with the redefined single-polymer activities
\beq
\tilde{z}_\sigma(\gamma) := z_\sigma(\gamma)\cdot (-1)^{\sigma\cdot\gamma},
\qquad(\gamma\in P_k,\ \sigma \in {\bf G}_k).
\Leq{redef}
By (\ref{eq:freie}) and (\ref{redef}) we get
\begin{eqnarray}
\lefteqn{\hat{Q}_k(t)- \delta_{t,0}\cdot \l(\log(2) +  k\log\l(\frac{4-r}{2}\ri)\ri)} \nonumber\\
&=& 2^{-k} \sum_{\sigma\in {\bf G}_k} \sum_{X\in C^{\infty,e}_k}
\frac{n(X)}{|X|!} \tilde{z}_\sigma^X\cdot  (-1)^{\sigma\cdot t} \nonumber \\
&=&  \sum_{X=(\gamma_1,\ldots,\gamma_{|X|})\in C^{\infty,e}_k} \frac{n(X)}{|X|!} z^X \cdot
2^{-k} \sum_{\sigma\in {\bf G}_k} (-1)^{\sigma\cdot (t+\sum_{i=1}^{|X|} \gamma_i)} \nonumber\\
&=& \sum_{\stackrel{X=(\gamma_1,\ldots,\gamma_{|X|})\in C^{\infty,e}_k}{\sum_{i=1}^{|X|} \gamma_i=t}}
\frac{n(X)}{|X|!} z^X,
\label{CCC}
\end{eqnarray}
using the identity $\sum_{\sigma\in {\bf G}_k} (-1)^{\sigma\cdot s} =
2^k \delta_{s,0}$.
As shown in \cite{GuK} as a consequence of Thm.\ 4, for the graph $G=(V,E)$
\[{\rm sign}(n(G)) = \l\{
\begin{array}{ll}
0 & ,\ G \mbox{ not connected}\\
-(-1)^{|V|} & ,\ G \mbox{connected}
\end{array}
\ri. .\] So, noticing the negative signs of the activities in
(\ref{activity}), all terms on the r.h.s.\ of (\ref{CCC}) are
nonpositive, proving the ferromagnetic property. \hfill $\Box$

\vskip 0.2cm
\noindent
We now rewrite the Fourier coefficients
$\hat{q}_k(t)\ (t\in \Gb_k)$ of Proposition \ref{prop:7.2}, using
the previously defined map $\psi_k:\Gb_k\to \Gb_k$,
\[\!\!\!\!\!\!\!\!\!\!\!
\psi_k(t_1,\ldots,t_k):=(t_1,t_1+t_2,\ldots,t_1+ \ldots+t_k) \quad
({\rm mod}\ 2).\] This is a group automorphism with inverse
\[\psi_k^{-1}(s_1,\ldots,s_k)=
(s_1,s_1+s_2,s_2+s_3,\ldots,s_{k-1}+s_k)\quad ({\rm mod}\ 2).\]
\begin{proposition}
For parameter values $r\in[0,2)$, $k\in\bN_0$ and $t\in\Gb_k$
\beq
\!\!\!\!\!\!
\hq_k(t)=(1+(-1)^{|t|})(-1)^{\LA t,\psi_k(t)\RA}
\exp\Big(c_0k+c_1|\psi_k(t)|+c_2\LA t,\psi_k(t)\RA\Big)
\Leq{Z}
with constants $c_0(r):=\ln\left(\frac{4-r}{2}\right)$,
$c_1(r):=\ln\left(\frac{2-r}{4-r}\right)$, $c_2(r):=\ln\left(\frac{r}{4-r}\right)$.
\end{proposition}
{\bf Proof.}
Our starting point is formula (\ref{D}):
\beq
\hq_k(t)=\left(1+(-1)^{|t|}\right)\left(\frac{4-r}{2}\right)^k\prod_{i=1}^{n(t)}z(\gamma_i)
\qquad (t\in G_k),
\Leq{X}
with the activities $z(\gamma_i)$ defined in (\ref{activity}).
\\[2mm]
$\bullet$
For $r<2$ and even $|t|$ we get $\sign(\hat{q}_k(t))=(-1)^{n(t)}$, since then
\[z(\gamma_i)
=-\frac{r}{2-r}\left(\frac{2-r}{4-r}\right)^{\supp(\gamma_i)}<0\]
(for odd $|t|$ the Fourier coefficient $\hq_k(t)$ vanishes anyhow).\\
As $(\psi_k(t))_i=1$ iff $\sum_{\ell=1}^{i-1}t_\ell$ is odd,
$t_i(\psi_k(t))_i= 1$ for every second $i$ with $t_i=1$. So \beq
n(t)=\LA t,\psi_k(t)\RA \qquad (t\in \Gb_k). \Leq{Y} $\bullet$
Returning to the assumption that $|t|$ is even, we note that
\[|\psi_k(t)|=\sum_{i=1}^{n(t)}(\supp(\gamma_i)-1)=-n(t)+\sum_{i=1}^{n(t)}
\supp(\gamma_i).\]
We now can write (\ref{X}) in the form
\beqno
\lefteqn{\hspace*{-10mm}\hq_k(t)= \Big(1+(-1)^{|t|}\Big)(-1)^{n(t)}}\\
&&\hspace*{-8mm}\exp\left(k\ln\left(\frac{4-r}{2}\right)+
\sum_{i=1}^{n(t)}
\ln\left(\frac{r}{2-r}\right)+\left(\sum_{i=1}^{n(t)}\supp(\gamma_i)\right)\ln\left(\frac{2-r}
{4-r}\right)\right).
\eeqno
Substituting (\ref{Y}), we obtain formula (\ref{Z}).
\hfill $\Box$
\begin{corollary}
For $k\in\bN$ and $t\in \Gb_k$ we have $\hq_k(t)=0$ for $|t|$ odd, and
for $|t|$ even, setting $\sigma_i:=(-1)^{(\psi_k(t))_i}$, $i=1,\ldots,k$
\[|\hq_k(t)|=c_k\exp\left(\tilde{c}_2\sum_{i=1}^{k-1}\sigma_i\sigma_{i+1}+\sum_{i=1}
^k\tilde{c}_1(i)\sigma_i\right),\]
with $\tilde{c}_2=\tilde{c}_1(1)=\tilde{c}_1(k)=-\frac{1}{4}\ln\left(\frac{4-r}{r}\right)<0$ and
$\tilde{c}_1(2)=\ldots=\tilde{c}_2(k-1)=\frac{1}{2}\ln\left(\frac{4-r)^3}{r(2-r)}\right)$.
\end{corollary}
{\bf Proof.}
We change from additive to multiplicative notation of the group elements,
that is, from $t_i\in\{0,1\}$ to $(-1)^{t_i}\in\{1,-1\}$.\\
Then the exponent in (\ref{Z}) can be written as linear combinations
of the terms $(-1)^{t_i+t_k}$, $(-1)^{t_i}$ and $t$--independent
constants. \\[2mm]
In terms of
$s:=\psi_k(t)$ we get, using $s_i=\eh(1-(-1)^{s_i})$
\beqno
\lefteqn{\hspace*{-10mm}
c_0k+c_1|\psi_k(t)|+c_2\LA t,\psi_k(t)\RA=
c_0k+c_1|s|+c_2\LA \psi_k^{-1}(s),s\RA}\\
&=&
{c}_0k+\frac{1}{2}\l(k-\sum_{i=1}^k(-1)^{s_i}\ri){c}_1\\
&&\hspace*{-8mm}+\frac{c_2}{4}\l(3k+1+
(-1)^{s_1}+(-1)^{s_k} -4\sum_{i=1}^{k}(-1)^{s_i}
+\sum_{i=1}^{k-1}(-1)^{s_i}(-1)^{s_{i+1}}\ri)\\
&=&
\tilde{c}_0(k)+\tilde{c}_1^{I}\Big( (-1)^{s_1}+(-1)^{s_k} \Big) +
\tilde{c}_1^{II}\sum_{i=1}^{k}(-1)^{s_i}
+\tilde{c}_2\sum_{i=1}^{k-1}(-1)^{s_i}(-1)^{s_{i+1}}
\eeqno
with $\tilde{c}_0(k):=k c_0+\eh k c_1+(3k-1)\frac{c_2}{4}$,
$\ \tilde{c}_1^{I}:=\frac{c_2}{4}$,
\[\tilde{c}_1^{II}:=-(\eh c_1+c_2)=
\frac{1}{2}\ln\l(\frac{(4-r)^3}{r(2-r)}\ri)\qmbox{ and }
\tilde{c}_2:=\frac{c_2}{4}=-\frac{1}{4}\ln
\l(\frac{4-r}{r}\ri)<0.\hspace*{6mm}\Box\]
\begin{remark}
In this sense $\hq_k(t)$ equals, up to a sign, the
Boltzmann factor of a 1D anti-ferromagnetic Ising system, whose two-body
interaction is of nearest neighbor form and translation invariant.

In the sense discussed in \cite{GuK}, the function $\hq_k$ equals the correlation
function of the spin system at inverse temperature $-1$, up to a normalisation
factor.
So the anti-ferromagnetic character (the negative sign of $\tilde{c}_2$) of the
interaction is due to negativity of the inverse temperature.

Several mathematicians, beginning with Kac (see his Comments in
P\'{o}lya \cite{Po}, pp.\ 424--426),
Newman \cite{Ne} and Ruelle \cite{Ru2}, conjectured the
existence of a Ising spin system related to the Riemann zeta function.
\\[2mm]
One  motivation for that conjecture is the
{\em Lee-Yang circle theorem} of
statistical mechanics. It states
that all zeroes of the partition function
\[Z(h):= \sum_{X\subset\Lambda} \exp(h|X|) \prod_{x\in X}\prod_{y\in \Lambda-X}
a_{xy}\]
of a ferromagnetic ($a_{xy}=a_{yx}\in[-1,1]$) Ising model occur at imaginary values of the external magnetic
field $h$ (see \cite{Ru2} for a proof).
\\[2mm]
The significance -- if any -- of the above corollary, remains to be clarified.
\end{remark}
%
\Section{Thermodynamic formalism}
\subsection{Partition function and transfer operator}
%
We now establish a direct correspondence between the partition
function $Z_n^C$ of the spin chain (along with some generalizations
of it) and the transfer operator of the map $F_r$. To this end we
first extend the tree $\tree (r)$ by considering the tree
${\tilde\tree} (r)$ having $1\over 1$ as root node and ${\tree} (r)$
as its left sub-tree starting at the second row. Each row is then
completed by reflecting the corresponding row of ${\tree} (r)$ w.r.t
the middle column and acting on each leaf with the transformation
${\hat S}_r$ defined in (\ref{Sr}).  Using the above terminology,
the $n$-th row $R_n$ of ${\tilde\tree} (r)$ is given for $n>1$ by
\be R_n:=  \biggl(\tree_n(r)\setminus \tree_{n-1}(r)\biggr) \cup
{\hat S}_r\biggl(\tree_n(r)\setminus \tree_{n-1}(r)\biggr) \ee

\begin{figure}[h]
\begin{center}
\includegraphics[width=8cm]{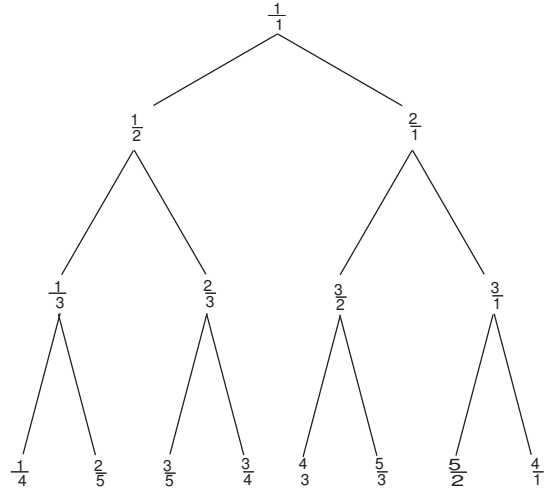}
\caption{{\small The Stern-Brocot tree.}}
\label{sternbrocot}
\end{center}
\end{figure}

\noindent
Note that ${\tilde\tree} (1)$ coincides with the classical
{\it Stern-Brocot tree} (see Fig.\ref{sternbrocot}).

\begin{proposition}\label{uno}For all $r\in [0,2)$, $x\in \R_+$, $s\in \C$ and $n \geq 1$ we have
 \be\label{complete} ({\cal
P}^n_{s,r} \,1 )(x)= 2\,\rho^{ns}\sum_{{p\over q}\in R_n }( p\,r\, x+ \rho q)^{-2s}. \ee
\end{proposition}
{\bf Proof.}
For $n=1$ we have
$$
({\Psr}1)(x)={2\rho^{s}\over (rx+\rho )^{2s}}.
$$
Suppose that (\ref{complete}) holds true. Then
$$
({\cal P}^{n+1}_{s,r} \,1) (x)= 2\,\rho^{(n+1)s}\sum_{{p\over q}\in
R_n }{1\over (\rho +rx)^{2s}}\left[{1\over ( {p r x\over \rho
+rx}+\rho\, q)^{2s}}+  {1\over ( pr -{p r x\over \rho +rx}+\rho\,
q)^{2s}}\right] $$
$$= 2\,\rho^{(n+1)s}\sum_{{p\over
q}\in R_n }\left[{1\over ( (p+\rho q) rx+\rho^2\, q)^{2s}}+  {1\over ( (p(r-1)+\rho q) rx+\rho(pr+\rho
q))^{2s}}\right]
$$
We now note that
$$
{ p(r-1)+\rho q \over pr+\rho  q}=\Phi_{1}\left({p\over q}\right)\quad\hbox{and}\quad
{ p+\rho q \over \rho  q}=S_r\left(\Phi_{1}\left({p\over
q}\right)\right).
$$
Therefore, if ${p\over q} \in R_n\cap \tree (r)$ then ${ p(r-1)+\rho
q \over pr+\rho q}\in R_{n+1}\cap \tree (r)$. On the other hand, if
${p\over q} \in R_n\cap ({\tilde\tree} (r) \setminus \tree (r))$,
then ${p'\over q'} = {\hat S}_r({p\over q}) \in R_n \cap\tree (r)$
and therefore $\Phi_{1}({p\over q})= \Phi_{0}({p'\over q'})$.
This allows to conclude that the last line in the above expression
is (\ref{complete}) with $n$ replaced by $n+1$  and the proof
follows by induction.
\hfill $\Box$

\begin{corollary}\label{coro}
For all $r\in [0,2)$, $k\in \N_0$ and $s\in \C$,
\be\label{figa} \sum_{\sigma \in {\bf G}_k}q_k^{-2s}(\sigma) = {1\over 2}\,\rho^{-(k+1)s}\, ({\cal P}_{s,r}^{k+1}1)(1)\ee
(with $\rho =2-r$) and therefore
\be\label{cano}
2\, Z^C_{n-1}(2s) = 1+\sum_{k=0}^{n-1}\rho^{-ks}\,(\Psr^k 1)(1).
\ee
\end{corollary}
The proof of this corollary follows at once from identity (\ref{complete}) along with the following lemma, whose proof
amounts to an elementary calculation.

\begin{lemma}\label{lemmino}  If
$$ \quad 1>{p\over q} \in \tree (r)$$
then
$$ 1<{p'\over q'}:={\hat S}_r\left({p\over
q}\right)={(r-1)p+(2-r)q \over rp +(1-r)q} \in {\tilde\tree}
(r)\setminus \tree (r)
$$
and
$$
r\, p'+\rho\, q'=r\, p+\rho\, q
$$
Therefore $r\, p'+\rho\, q'$ is the denominator of both descendants of ${p\over q}$ in $\tree (r)$.
\end{lemma}
%

\subsection{Phase transitions}
%
If we define a 'grand-canonical' ensemble as in \cite{Kn3} where the
partition function is given for $k \in\bN_0$ by \be \label{ZG}
Z_k^G(s):=\sum_{\sigma \in {\bf G}_k}\exp{(-s\, Q_k(\sigma))}\equiv
\sum_{{p\over q}\in \tree_{k+1}(r)\setminus\tree_{k}(r)}q^{-s},\quad
k\geq 0 \ee then \beq \Zc(s)= 1+\sum_{k=0}^{n-1}Z^G_k(s),\qquad
n\in\bN. \Leq{Zc:Zn} Due to Corollary \ref{coro}, the existence of a
spectral gap for all $r\in [0,1)$ and standard arguments of
thermodynamic formalism \cite{Ru1} all 'grand canonical'
thermodynamic functions are analytic for all $s\in \C$ and there is
no phase transition. On the other hand, in the canonical setting,
using (\ref{cano}) and observing that the denominators $q_k(\sigma)$
are monotonically decreasing functions of $r$ the limit  $\lim_{n\to
\infty} Z^C_n(s)$ exists and is finite for $\Re \, s$ large enough.
This suggests that in this framework there is indeed a phase
transition. More specifically, let
$$
F_n(s,r):=\frac{1}{n}\log\left(2Z^C_{n-1}(2s)\right),\qquad
n\in\N,\quad s\in\R,\quad r\in [0,1],
$$
and \beq F(s,r):= \lim_{n\to\infty} F_n(s,r). \Leq{freelimit} In
terms of the canonical expectations $\left<\,\cdot
\,\right>_{n,s,r}$ the {\it mean magnetization} is defined as \beq
M(s,r):=\lim_{n\to\infty} M_n(s,r), \qmbox{where} M_n(s,r):=
\left<\frac{1}{n}\sum_{k=1}^n(-1)^{\sigma_k}\right>_{n,s,r}.
\Leq{magneto}
\begin{remark}
We remind the reader of the notions of {\it order of a phase
transition}: for a free energy density $F:\R\to\R$, $s\in\R$ is
called a {\em phase transition of order} $n\in\N$ if $F$ is $n-1$ times,
but not $n$ times continuously differentiable at $s$.
\end{remark}
\begin{theorem}\label{fase}
\begin{enumerate}
\item
The limit in (\ref{freelimit}) exists and $F\in C(\R\times [0,1])$.
\item Set
\beq \lambda_{s}\equiv\lambda_{s,r}:={\rm spec}\, {\rm rad}\left(\Psr\right)
\Leq{def:lambda} Then the function $s_{cr}:[0,1]\to\R$ of $r$, defined as
the smallest positive real solution of the equation 
\be \label{scr}
\lambda_{s/2,r} = (2-r)^{s/2}\ee 
is real analytic, increasing and
convex, with $s_{cr}(0)=1$ and $s_{cr}(1)=2$.
\item
For $r\in [0,1]$ there is a phase transition at $s=s_{cr}(r)$. More
precisely, for each $r\in [0,1]$,
 $s\mapsto F(s,r)$ is real analytic on $\R\setminus \{s_{cr}(r)\}$
 and
 $F(s,r)>0$ for $s < s_{cr}(r)$ whereas $F(s,r)=0$ for $s\geq s_{cr}(r)$.

\noindent If $0\leq r< 1$, the phase transition is of first order,
whereas  if $r=1$, it is of order two.
\item
The limit (\ref{magneto}) exists for all $s\in\R$, $r\in [0,2)$ and
\be M(s,r)=\begin{cases} 0, & s<s_{cr}(r), \cr 1, & s>s_{cr}(r). \cr
\end{cases} \ee
\end{enumerate}
\end{theorem}

{\bf Proof:}
\begin{enumerate}
\item
To prove the first assertion, note that since $s$ is real, eq. (\ref{cano}) yields
$$
\exp\left(F_n(s,r)\right)\,=\, \left(1+\sum_{k=0}^{n-1}\rho^{-ks}\,(\Psr^k 1)(1)\right)^{1/n}.
$$
The existence of the limit for $r<1$ is a simple consequence of the
spectral decomposition of the transfer operator $\Psr:{\cal
H}_s\to{\cal H}_s$, $r\in [0,1), s\in\R$, \beq \Psr^k f\,=\,
\lambda_s^k\, h_s\,\nu_s(f) + {\cal N}_s^k \,f,\qquad k\geq 1
\Leq{deco} where
$$
\Psr \, h_s=\lambda_s\, h_s, \quad \Psr^* \, \nu_s=\lambda_s\, \nu_s
$$
and $\lambda_s^{-k}\parallel {\cal N}_s^k\parallel \to 0$,
$k\to\infty$. For $r=1$ the existence of the limit has been shown in
\cite{Kn4}. The continuity follows from standard convexity
arguments.
\item
By eventually passing to the common (for different $s$ values)
Banach space $H_{\infty}(D_1)$ of functions analytic in the disk
$D_1$, (see (\ref{disk})) and continuous on $\partial D_1$, the
family of operators $\Psr:H_{\infty} \to H_{\infty}$ becomes an
analytic family of type A in the sense of Kato (\cite{Ka}, Chapter
7). Therefore, for $s\in\R$, $s<s_{cr}(r)$, the function $\lambda_s$
defined in (\ref{def:lambda}) is real analytic and monotonically
decreasing.

The other claimed properties of the function $s_{cr}(r)$ can be
easily obtained by differentiating twice the logarithm of
(\ref{scr}).
\item
The first statement of point (3) of the theorem is now a direct
consequence of what just proved and the identity (\ref{cano}).

The proof that the phase transition for $r=1$ is of second order is contained
in \cite{CK} and \cite{PS}.

So let $r<1$ and  $\delta:=s_{cr}-s\geq 0$. The real analytic function
$g(\delta):= \rho^{-s}\lambda_s$ is convex and increasing in $\delta$, with $g(0)=1$.

\noindent Using once more (\ref{deco}) and setting $a_s:=h_s(1)
\nu_s(1)$, we can write \be \sum_{k=0}^{n-1}\rho^{-ks}\,(\Psr^k
1)(1)\,=\, \frac{g^n(\delta)-1}{g(\delta)-1} \left(a_s\,+\,
R_n(\delta)\right), \ee with $R_n(\delta)={\cal O}(1)$ in both
limits $\delta \to 0$ and $n\to \infty$. Therefore, using the
expansion $g(\delta)=1+a\delta+o(\delta)$, $a>0$, we get
$F(s,r)=a\delta + o(\delta)$ as $\delta\to 0$, which is what we
needed.
\item
Concerning the last statement, note first that from the identity
(\ref{simmetria}) it follows that the "grand canonical" energy
function $ Q_k=\log q_k :{\bf G}_k\to\R$ has the symmetry
$Q_k(\sigma)=Q_k({\bar\sigma})$. Moreover, if we denote with
$0_k=(0,0,\ldots,0)\in {{\bf G}_k}$, it is immediate to see that for
an arbitrary function $f$ on ${{\bf G}_n}$, we have
$$
\sum_{\sigma\in {\bf G}_n} f(\sigma)\,=\, f(0_n) + f(10_{n-1})
+\sum_{m=1}^{n-1}\sum_{\tau\in {\bf G}_m} f(\tau,1,0_{n-m-1}).
$$

Therefore putting these observations together with (\ref{cano2}) and
(\ref{ZG}) we can write
\begin{eqnarray}
\lefteqn{\Zc(s)M_n(s,r) = \sum_{\sigma\in {{\bf G}_n}}\left(\frac{1}{n}\sum_{k=1}^n (-1)^{\sigma_k}\right)
\left(q_n^c(\sigma)\right)^{-s}}\NN\\
&=& 1+(1-\frac{2}{n})2^{-s} + \sum_{m=1}^{n-1}\sum_{\tau\in{\bf G}_m}\frac{n-m-2+
\sum_{j=1}^m (-1)^{\tau_j}}{n}\left(q_m(\tau)\right)^{-s} \NN\\
&=& 1+(1-\frac{2}{n})2^{-s} + \sum_{m=1}^{n-1}\frac{n-m-2}{n} Z^G_m(s)\NN\\
&=& 1+\sum_{m=0}^{n-1}\frac{n-m-2}{n} Z^G_m(s).
\label{prev:id}
\end{eqnarray}
If $s>s_{cr}$, then the limit $\lim_{n\to\infty} \Zc(s)$ is finite and
the factors $(n-m-2)/{n}$ in (\ref{prev:id}) go to one.
Comparison of (\ref{Zc:Zn}) with (\ref{prev:id})
immediately implies that
$\lim_{n\to\infty} M_n(s,r)$ exists and equals $1$.

If instead $s<s_{cr}$, then the ferromagnetic property (Thm.\
\ref{thm:ferro}) implies \beq M_n(s,r)\ge 0. \Leq{Mn:g0} We
follow a similar strategy as in the proof of Lemma 9 of \cite{CK},
but now based on the transfer operator analysis. By (\ref{deco}) and
(\ref{def:lambda}) for $2s<s_{cr}$ and $\vep\in\Big(0,
\eh(1-\rho^s/\lambda_s)\Big)$ (which is a nonempty interval since 
$s\mapsto \rho^s/\lambda_s$ monotonically increases to 1 for $2s=s_{cr}$)
\beq
\mu_\pm:= (1\pm\vep)\frac{\lambda_s}{\rho^s}>1.
\Leq{def:mu}
Using again the spectral decomposition (\ref{deco}) of the transfer
operator and Corollary \ref{coro},
\begin{eqnarray*}
\lefteqn{2 \l(\Zg(2s) - \mu^{n-l}_\pm Z^G_l(2s)\ri)}\\
&=& \rho^{-(n+1)s}\Big(a_s\lambda_s^{n+1} +({\cal
N}_s^{n+1}1)(1)\Big)-
\mu_\pm^{n-l}\rho^{-(l+1)s}\l(a_s\lambda_s^{l+1} +({\cal N}_s^{l+1}1)(1)\ri)\\
&=& \l(\frac{\lambda_s}{\rho^s}\ri)^{n+1}\l(a_s+\frac{({\cal
N}_s^{n+1}1)(1)}{\lambda_s^{n+1}}-
\l(\frac{\mu_\pm\rho^{s}}{\lambda_s}\ri)^{n-l}
\l(a_s +\frac{({\cal N}_s^{l+1}1)(1)}{\lambda_s^{l+1}}\ri)\ri)\\
&=& \l(\frac{\lambda_s}{\rho^s}\ri)^{n+1} \l(a_s+\frac{({\cal
N}_s^{n+1}1)(1)}{\lambda_s^{n+1}}- \l(1\pm\vep\ri)^{n-l}\l(a_s
+\frac{({\cal N}_s^{l+1}1)(1)}{\lambda_s^{l+1}}\ri)\ri).
\end{eqnarray*}
There exist $\delta\in(0,1)$ and $C\ge1$ with
\[\l|\frac{({\cal N}_s^{k}1)(1)}{\lambda_s^{k}}\ri|\le
{ \l\|{\cal N}_s^{k}\ri\| \over \lambda_s^{k}} \le C\delta^k\qquad
,\ k\in\bN.\] Thus there exists a $n_{\rm min}$ such that for all
$n\geq n_{\rm min}$ and for all $l={0,\ldots,n}$
\[2 \l(\Zg(2s) - \mu^{n-l}_- Z^G_l(2s)\ri) \ge
\l(\frac{\lambda_s}{\rho^s}\ri)^{(n+1)} \l(a_s-
\l(1-\vep\ri)^{n-l}\l( a_s +C\delta^{l+1}\ri)\ri) \ >0\] and
similarly
\[2 \l(\Zg(2s) - \mu^{n-l}_+ Z^G_l(2s)\ri) \le
\l(\frac{\lambda_s}{\rho^s}\ri)^{(n+1)} \l( a_s+C\delta^{n+1} -
 a_s\l(1+\vep\ri)^{n-l}\ri)<0.\]

\noindent In other words, rescaling $2s\to s$ and the constants
$\mu_\pm$ accordingly, we have for all $s< s_{cr}$
 \beq \mu_+^{l-n}
Z^G_n(s) \leq Z^G_l(s) \leq \mu_-^{l-n} Z^G_n(s), \qquad
l\in\{0,\ldots,n\}. \Leq{ineq:ZZZ}

So we have for $n\geq n_{\rm min}$, with inequality  (\ref{def:mu})
\beq
\hspace*{-4mm}
\Zc(s) = 1 + \sum_{l=0}^{n-1} Z^G_l(s) \geq
\l( \sum_{l=0}^{n-1} \mu_+^{l-n+1}  \ri) Z^G_{n-1}(s)=\frac{1-\mu_+^{-n}}{1-\mu_+^{-1}} Z^G_{n-1}(s).
\Leq{ineq:CG}

Now we use the upper bound in (\ref{ineq:ZZZ})
for the grand canonical partition function $Z^G_l$.
\begin{eqnarray*}
\lefteqn{\Zc(s) M_n(s,r)
= 1 + \sum_{m=0}^{n-1} \frac{n-m-2}{n} Z^G_m(s)
= 1 + \sum_{l=0}^{n-1} \frac{l-1}{n} Z^G_{n-1-l}(s)} \\
& \leq & 1 + Z^G_{n-1}(s) \cdot \sum_{l=0}^{n-1}\frac{l-1}{n} \mu_-^{-l}
= 1 + \frac{Z^G_{n-1}(s)}{n} \cdot\frac{d}{d \mu_-} \l(-\sum_{l=0}^{n-1}\mu_-^{-l+1} \ri) \\
&=& 1 + \frac{Z^G_{n-1}(s)}{n} \cdot\frac{d}{d \mu_-}
\frac{\mu_{-} -\mu_-^{-n+1}}{1-\mu_-^{-1}} \\
&=& 1 + Z^G_{n-1}(s) \l[ \frac{\mu_-^{-n+1}}{\mu_- - 1} +
\frac{1}{n} \frac{(1 - \mu_-^{-n})(1-2\mu_-^{-1})}{(1-\mu_-^{-1})^2} \ri].
\end{eqnarray*}
With the lower bound (\ref{ineq:CG}) for $\Zc(s)$ we obtain
\[M_n(s,r) \leq \l( (Z^G_{n-1}(s))^{-1} + \frac{\mu_-^{-n-1}}{(\mu_- - 1)}
+ \frac{1}{n} \frac{1}{(1-\mu_-^{-1})^2} \ri)
\l/ \l( \frac{1-\mu_+^{-n}}{1-\mu_+^{-1}}   \ri) \ri. ;\]
since $\lim_{n\ar\infty} Z^G_{n-1}(s)=\infty$, and
$\mu_- > 1$, this implies
$$\limsup_{n\ar\infty} M_n(s,r) \leq 0.$$
Together with (\ref{Mn:g0}) we see that the limit in (\ref{magneto})
exists and equals $0$.\hfill $\Box$
\end{enumerate}
\vskip 0.5cm
\begin{remark}
Note that (only) for $r=1$ (due to arithmetical quibbles) we have
$$
{\cal P}_{s,1}^n1(0) = 1+\sum_{k=0}^{n-1}{\cal P}_{s,1}^k1(1)
$$
and therefore
$$
2\, Z_{n-1}^C(2s)={\cal P}_{s,1}^n1(0)
$$
This makes the `canonical' and `grand canonical' settings equivalent
at all temperatures for $r=1$. But for $r\ne 1$ this equivalence
fails below $s_{cr}^{-1}$.
\end{remark}

\begin{example}\label{zero}
For $r=0$ we find
$$
Z^C_n(s)={2^s-1-2^{n(1-s)}\over 2^s-2}\quad\hbox{so that}\quad
\lim_{ n\to \infty} Z^C_n(s) = {2^s-1\over 2^s-2}
$$
for $\Re\, s>1$ (equation (\ref{scr}) becomes $2^{1-{s\over
2}}=2^{s\over 2}$). Fore real $s$ the free energy is given by
$$
F(s,0)=\lim_{n\to \infty}{1\over n} \log Z^C_n(s) =
\begin{cases}
(1-s)\log 2, &  \; s < 1 \cr
0, & \; s\geq 1. \cr
\end{cases}
$$
\end{example}
\vskip 0.5cm
\begin{example}
For $r=1$ one finds \cite{Kn1} $$\lim_{n\to \infty}
Z_n^C(s)=\sum_{n\geq 1}{\varphi(n)\over n^s}={\zeta (s-1)\over \zeta
(s)}, \qquad \Re\, s > 2, $$ where $\zeta (s)$ is the Riemann zeta
function. Moreover one can show that
$$
Z_{n}^C(2) \sim {n\over 2\log n}, \quad n\to \infty.
$$
The free energy $F(s,1)$ is real analytic for $s <2$ and \cite{PS}
$$
F(s,1) \sim {2-s \over -\log{(2-s)}} \quad \hbox{as}\quad s\nearrow 2.
$$

\end{example}
%

\subsection{Fourier analysis of the transfer operator}

\vskip 0.5cm
\noindent
Up to now we mainly analysed the action of the transfer operator on
positive functions, related to the Perron-Frobenius eigenfunction.
Now we are interested in the spectral gap and its disappearance
for $r\nearrow 1$.
Therefore we extend Proposition \ref{uno} by applying $\Psr^n$ to
$e_m(x):=e^{2\pi im x}$.
\begin{proposition}\label{caratteri}
For all $r\in [0,2)$, $x\in \R_+$, $s\in \C$, $m\in \Z$ and $n \geq 1$ we have
\be \label{formulona1}
\l(\Psr^n e_m\ri)(x)=
\rho^{ns}\, \sum_{{p\over q}\in R_n }
\frac{
e_m\,\l({n_{0}(x,\,p/q)\over  pr x+\rho q}\ri)
+e_m \,\l({n_{1}(x,\, p/q)\over  pr x+
\rho q}\ri)}
{( pr x+
\rho q)^{2s}}
\ee
where the functions $n_0$ and $n_1$ satisfy:
\be\label{prop1}
n_{0}(x,\, p/q) +n_1(x,\, p/q)= p\,r\, x+
\rho \,q,\quad \forall x \in \R_+.
\ee
More specifically,
\begin{eqnarray}
n_0(x,\,p/q)&=&\mu\, x+\rho \,\nu  \label{w0}\\
n_1(x,\,p/q)&=&(pr-\mu)x+\rho (q-\nu) \label{w1}
\end{eqnarray}
for some choice of numbers
 $0\leq \mu \leq pr$ and $0\leq \nu \leq q$.
\end{proposition}
{\bf  Proof.}
For $n=1$ we find
$$
\l(\Psr e_m\ri)(x)=\rho^{s}\,
\frac{e_m\l({x\over rx+\rho}\ri)+
e_m\l(\frac{(r-1)x+\rho}{rx+\rho}\ri)}{(rx+\rho )^{2s}}
$$
whereas for $n=2$
\begin{eqnarray}
\l(\Psr^2 e_m\ri)(x)&=&\rho^{2s}\,
\left[
\frac{e_m\left({x+\rho \over rx+2\rho}\right)+
e_m\left({(r-1)x+\rho\over rx+2\rho}\right)}
{ (rx+2\rho\ )^{2s}}  +\right. \nonumber \\
&&\qquad\left. \frac{e_m\left({x \over (3-r)rx+\rho^2}\right)
+e_m\left(\frac{((3-r)r-1)x+\rho^2}{(3-r)rx+\rho^2}\right)}{ ((3-r)rx+\rho^2\ )^{2s} }
\right] \nonumber
\end{eqnarray}
Hence formula (\ref{formulona1}) holds for $n=1$ with the choice $\mu=1, \nu=0$ and for $n=2$ with the choice $\mu=1,
\nu=1$. If we set
$$
V_i(x,{p\over q},\mu,\nu) := {n_{i}(x,\,p/q)\over  pr x+ \rho q},\qquad i=0,1,
$$
we find, for $i=0,1$,
$$
V_{i}(\Phi_{0}(x),{p\over q},\mu,\nu)=V_{i}(x,{p+\rho q\over \rho q},\mu+r\rho \nu,\rho \nu),
$$
and
$$
V_{i}(\Phi_{1}(x),{p\over q},\mu,\nu)=V_{i}(x,{p(r-1)+\rho q\over
pr+\rho q},\mu(r-1)+\nu, \mu+ \nu).
$$
The proof now proceeds by induction along the same lines as for Proposition \ref{uno}. \hfill $\Box$

\vskip 0.5cm
\noindent
A more precise characterization of the integers $\mu$ and $\nu$ appearing in the definition of the functions $n_0$ and $n_1$
would be of some interest (see however below, Proposition \ref{andreaslemma}). Nevertheless, property (\ref{prop1}), along with Lemma
\ref{lemmino},  is sufficient to have the following extension of Corollary
\ref{coro}:
\begin{corollary} For all $r\in [0,2)$, $k\in \N_0$,  $s\in \C$ and $m\in \Z$
\be
\sum_{\sigma \in {\bf G}_k}
q_k^{-2s}(\sigma)\, e_m\left({p_k(\sigma)\over q_k(\sigma)}\right)
= \eh\,\rho^{-(k+1)s}\, \left({\cal P}_{s,r}^{k+1}e_m\right)(1).
\ee
\end{corollary}
The proof of Proposition \ref{formulona1} extends at once to
arbitrary complex functions $f:[0,1]\to\bC$. On the other hand we
shall formulate and prove this more general result in a direct way,
without employing the extended tree ${\tilde\tree} (r)$.
\begin{proposition}\label{andreaslemma}
For all $r\in[0,2),\ k\in\bN_0,\ s\in\bC$ and $f:[0,1]\to\bC$
\beqn
\lefteqn{\eh\rho^{-(k+1)s}\l(\cP_{s,r}^{k+1}f\ri)(x)=
\sum_{\sigma\in {\bf G}_k}
\big(q_k(\sigma)-(1-x)rp_k(1-\sigma)\big)^{-2s}.}\nonumber\\
&&\eh\sum_{i=0,1}f\l(\frac{p_k(\sigma)-(1-x)s_{k+1}(\sigma,i)}
{q_k(\sigma)-(1-x)t_{k+1}(\sigma,i)}\ri) \quad (x\in[0,1]),
\label{it}
\eeqn
with functions $s_k\equiv s_{k,r}$ and
$t_k\equiv t_{k,r} :{\bf G}_k\ar\bR$ defined by
$s_0:=1,\ t_0:=0$,
\beqno
s_{k+1}(0,\sigma)&:=&s_k(\sigma), \\
s_{k+1}(1,\sigma)&:=&(r-1)s_k(\overline{\sigma})+(2-r)t_k(\overline{\sigma}),\\
t_{k+1}(0,\sigma)&:=&rs_k(\sigma)+(2-r)t_k(\sigma), \\
t_{k+1}(1,\sigma) &:=& rs_k(\overline{\sigma})+(2-r)t_k(\overline{\sigma}).
\eeqno
\end{proposition}
{\bf Proof.} By definition the transfer operator acts like \beq
(\cP_{s,r}f)(x)=
\frac{\rho^s}{(\rho+rx)^{2s}}\big[f(\Phi_{0}(x))+f(\Phi_{1}(x))\big]
\qquad (x\in[0,1]), \Leq{traop} with $\rho=2-r$. It is well--defined
for the range $0\leq r<2$.
\\[2mm]
$\bullet$
For $k=0$ we have as arguments of $f$ in (\ref{it})
\[\Phi_{0}(x)=\frac{1-(1-x)}{2-r(1-x)} \qmbox{and} \Phi_{1}(x)=\frac
{1-(r-1)(1-x)}{2-r(1-x)}\]
which in this case gives the formula, since
\[s_1(0)=1\qmbox{,}s_1(1)=r-1 \qmbox{and} t_1(0)=t_1(1)=r.\]

$\bullet$
For $k\in \bN_0$
\beqno
\lefteqn{\Phi_{0}\l(
\frac   {p_k(\sigma)-(1-x)s_{k+1}(\sigma,i)}
      {q_k(\sigma)-(1-x)t_{k+1}(\sigma,i)}\ri)}\\
&=& \frac{[(2-r)q_k(\sigma)+(r-1)p_k(\sigma)] -(1-x) [s_{k+1}(\sigma,i)]}
{[(2-r)q_k(\overline{\sigma})+rp_k(\overline{\sigma})]
-(1-x) [r s_{k+1}(\overline{\sigma},i)-(2-r)t_{k+1}(\overline{\sigma},i)]}\\
&=& \frac{q_{k+1}(0,\sigma)-(1-x)s_{k+2}(0,\sigma,i)}
{p_{k+1}(0,\sigma)-(1-x)t_{k+2}(0,\sigma,i)}
\eeqno
and, with $i':=1-i$,
\beqno
\lefteqn{\hspace*{-10mm}\Phi_{1}(x)\l(\frac{p_k(\overline{\sigma})-(1-x)s_{k+1}(\overline{\sigma},i')}
 {q_k(\overline{\sigma})-(1-x)t_{k+1}(\overline{\sigma},i')}\ri)}\\
&\hspace*{-22mm}=& \hspace*{-12mm}
\frac{[(2-r)q_k(\overline{\sigma})+(r-1)p_k(\overline{\sigma})] -(1-x) [(r-1)s_{k+1}(\overline{\sigma},i')+(2-r)t_{k+1}(\overline{\sigma},i')]}
{[(2-r)q_k(\overline{\sigma})+r p_k(\overline{\sigma})] -(1-x) [r s_{k+1}(\overline{\sigma},i')-(2-r)t_{k+1}(\overline{\sigma},i')]}\\
&\hspace*{-22mm}=& \hspace*{-12mm}\frac{q_{k+1}(1,\sigma)-(1-x)s_{k+2}(1,\sigma,i)}{p_{k+1}(1,\sigma)-(1-x)t_{k+2}(1,\sigma,i)}
.
\eeqno
$\bullet$
Concerning the $(k+1)$--th iterates of the factor $\frac{\rho^s}{(\rho+rx)^{2s}}$
in (\ref{traop}), the induction step reads,
substituting first $p_k(\overline{\sigma}) = q_k(\sigma)-p_k(\sigma)$
\beqno
\lefteqn{\frac{\rho^s}{(\rho+rx)^{2s}}
 \l( \big( q_k(\sigma)-(1-\Phi_{0}(x))rp_k(\overline{\sigma}) \big)^{-2s}\ri)}\\
&=& \rho^s\big( \big[(2-r)q_k(\sigma)+r p_k(\sigma)\big]
-(1-x)r [(2-r)q_k(\sigma)+(r-1)p_k(\sigma) \big)^{-2s}\\
&=& \rho^s\big(
q_{k+1}(0,\sigma)-(1-x)rp_{k+1}(1,\overline{\sigma})\big)^{-2s}
\eeqno since $q_{k+1}(0,\sigma)=(2-r)q_k(\sigma)+r p_k(\sigma)$ and
$p_{k+1}(1,\overline{\sigma})=(2-r)q_k(\sigma)+(r-1) p_k(\sigma)$.
Similarly \beqno \lefteqn{\frac{\rho^s}{(\rho+rx)^{2s}}
 \l( \big( q_k(\sigma)-(1-\Phi_{1}(x))rp_k(\overline{\sigma}) \big)^{-2s}\ri)}\\
&=& \rho^s\big( \big[(2-r)q_k(\sigma)+r p_k(\sigma)\big]
-(1-x)r p_k(\sigma) \big)^{-2s}\\
&=& \rho^s\big( q_{k+1}(1,\overline{\sigma})-(1-x)rp_{k+1}(0,\sigma)\big)^{-2s}
\eeqno
since  $q_{k+1}(1,\overline{\sigma}) = (2-r)q_k(\sigma)+r p_k(\sigma)$, too and
$p_{k+1}(0,\sigma)=p_k(\sigma)$.\hfill
$\Box$

\begin{corollary}
For all $r\in[0,2),\ k\in\bN_0,\ s\in\bC$ and $f:[0,1]\to\bC$
\[\eh\rho^{-(k+1)s}\l(\cP_{s,r}^{k+1}f\ri)(1)=\sum_{\sigma\in {\bf G}_k}
(q_k(\sigma))^{-2s}f\l(\frac{p_k(\sigma)}{q_k(\sigma)}\ri) .\]

\end{corollary}

\subsection{Twisted zeta functions}

For $m\in\bZ$ define: \be Z_n^{(m)}(s) := \sum_{{p\over q}\in
\tree_{n}(r)\setminus \{0\}}q^{-s} \, e^{2 \pi i \,m\, {p\over q}}.
\ee Then by the above \be 2\,Z_n^{(m)}(2s) =
1+\sum_{k=0}^{n}\,\rho^{-ks}\, \left({\cal
P}_{s,r}^{k}e_m\right)(1). \ee
\vskip 0.5cm
\begin{example}
For $m=r=1$ we have for $\Re \, s >2$
$$
\lim_{n\to \infty}Z_n^{(1)}(s)=\sum_{q\geq 1} {\mu (q) \over q^s} =
{1\over \zeta(s)}
$$
 since the M\"obius function
$$
\mu (\prod p^{n_p})=
\begin{cases}
(-1)^{\sum n_p}& , n_p\leq 1 , \cr 0&, {\rm otherwise} \,, \cr
\end{cases}
$$
satisfies
$$
\mu (q)= \sum_{0<p\leq q \atop gcd(p,q)=1}  e^{2 \pi i \, {p\over
q}}, \quad q\in \N.
$$
\end{example}
\vskip 0.5cm
\begin{remark} Defining $\zeta_0(s):=2^s/(2^s-1)$ from Example \ref{zero} we have
\be \lim_{n\to \infty} Z_n^C(s)={\zeta_0(s-1)\over \zeta_0(s)},
\qquad \Re\, s > 1. \ee One could guess that for all $r\in [0,1]$ it
holds \be\label{con1} \lim_{n\to \infty} Z^{C}_n(s)={\zeta_r
(s-1)\over \zeta_r (s)}, \qquad \Re\, s > s_{cr}, \ee with
$\zeta_1(s)=\zeta(s)$ and more generally, for $r\in [0,1]$,
$$
{1\over \zeta_r(s)}:=\lim_{n\to \infty}Z_n^{(1)}(s), \qquad \Re\,  s
> s_{cr}.
$$
On the other hand, a simple direct verification shows that for $r\ne
0,1$ this is not the case.
 Note that by the above we have
$\Re \,s > s_{cr}$ \be {2\over \zeta_r (2s)} =
1+\sum_{k=0}^{\infty}\,\rho^{-ks}\, \left({\cal
P}_{s,r}^{k}e_1\right)(1). \ee
\end{remark}
\vskip 0.5cm
 \noindent
 We conclude by discussing further $\zeta$'s for general
integer values of $m$ restricting to the case $r=1$.

\noindent
For real $x$ we set $e_x:\bC\to\bC,\ e_x(c):=e^{2\pi ixc}$.
The multiplicative group of units in the ring $\bZ/q\bZ$ is denoted by
$U(\bZ/q\bZ)$. It is of cardinality $\vv(q)$.
We are interested in the functions
\[\mu^{(m)}:\bN\to\bC \qmbox{,} \mu^{(m)}(q):=\sum_{p\in U(\bZ/q\bZ)}e_{m/q}(p) \qquad
(m\in\bZ,q\in\bN).\]
\begin{lemma}\label{5.11}
Denoting by $\mu$ the M\"obius function,
\[\mu^{(m)}(q)=\frac{\vv(q)}{\vv\l(\frac{q}{gc d(m,q)}\ri)}\mu\l(\frac
{q}{gc d(m,q)}\ri) \qquad (m\in\bZ,q\in\bN).\]
\end{lemma}
\begin{remark}
In particular $\mu^{(m)}$ is integer--valued, multiplicative,
\[\mu^{(-m)}=\mu^{(m)} \qmbox{,} \mu^{(0)}=\vv \qmbox{and} \mu^{(1)}=\mu.\]
\end{remark}
{\bf Proof.}
We set $q':=q/\gcd(m,q),\ m':=m/\gcd(m,q)$. Then
\beqno
\mu_m(q)&=&\sum_{p\in U(\bZ/q\bZ)}e_{m'/q'}(p)
=\frac{\vv(q)}{\vv(q')}\sum_{p'\in U(\bZ/q'\bZ)}e_{m'/q'}(p')\\
&=&\frac{\vv(q)}{\vv(q')}\sum_{p'\in U(\bZ/q'\bZ)}e_{1/q'}(p')
=\frac{\vv(q)}{\vv(q')}\mu(q').
\eeqno
The third equality is due to the fact that for $ m'$ relatively prime
to $q'$ multiplication by $m'$ only permutes the elements of $U(\bZ/q'\bZ)$.
\hfill $\Box$\\[2mm]
Next we consider the Dirichlet series
\[\zeta^{(m)}(s):=\sum_{n=1}^\infty\mu^{(m)}(n)n^{-s}\]
As $|\mu^{(m)}(n)|\leq n$, we know that these series converges
absolutely for $\Re(s)>2$. This is in fact also the abscissa of
unconditional convergence if $m$=0.
\begin{proposition}
For $m\in\bZ\backslash\{0\}$ the Dirichlet series $\zeta^{(m)}(s)$ converges
absolutely for $\Re(s)>1$.
\end{proposition}
{\bf Proof.}
This follows from Lemma \ref{5.11} and the estimate
\beq
\frac{\vv(q)}{\vv\l(\frac{q}{\gcd(m,q)}\ri)}\leq m
\Leq{test}
(\ref{test}) can be proven by taking the prime powers $p^a$ for $m$, since $\vv$ is multiplicative.
In that case we can also assume that $q=p^b$, and the inequality follows from
$\vv(p^c)=\vv(p)\, p^{c-1}$, valid for $c\geq1$.
\hfill $\Box$
%
%

\end{document}